\documentclass[aps,prb,reprint,amsmath,amssymb,graphicx,longbibliography]{revtex4-1}

\usepackage{bm}
\usepackage{dcolumn}
\usepackage{graphicx,subfigure}
\usepackage{bm}
\usepackage{verbatim}
\usepackage{amsmath}
\usepackage{amssymb}
\usepackage[T1]{fontenc}
\usepackage{ae,aecompl}
\usepackage{appendix}
\usepackage{float}

\newcommand{\cH}{{\mathcal H}}
\newcommand{\cL}{{\mathcal L}}

\newcommand{\bx}{{\bf x}}

\newcommand{\be}{\begin{equation}}
\newcommand{\ee}{\end{equation}}
\newcommand{\la}{\label}
\newcommand{\bea}{\begin{eqnarray}}
\newcommand{\eea}{\end{eqnarray}}
\newcommand{\p}{\partial}
\newcommand{\tr}{\text{tr}}

\newcommand{\pt}{{\partial}}

\newcommand{\rv}{{\bf r}}

\newcommand{\av}{{\bf a}}

\newcommand{\A}{{\bf A}}
\newcommand{\B}{{\bf B}}

\newcommand{\E}{{\bf E}}

\newcommand{\xv}{{\bf x}}
\newcommand{\yv}{{\bf y}}

\newcommand{\eps}{{\varepsilon}}
\newcommand{\bv}{{\bf b}}
\newcommand{\dv}{{\bf d}}
\newcommand{\ev}{{\bf e}}

\newcommand{\jv}{{\bf j}}
\newcommand{\tv}{{\bm\tau}}

\newcommand{\zh}{{\hat{\bf z}}}

\newcommand{\pv}{{\bf p}}

\newcommand{\oh}{{\frac{1}{2}}}

\newcommand{\grad}{{\bm{\nabla}}}

\newcommand{\vsigma}{{\bm{\sigma}}}
\newcommand{\vtau}{{\bm{\tau}}}

\newcommand{\bse}{\begin{subequations}}
\newcommand{\ese}{\end{subequations}}
\def\rf#1{(\ref{#1})}

\usepackage{color}

\definecolor{cardinal}{rgb}{0.6,0,0}
\definecolor{darkgreen}{rgb}{0,0.4,0}
\definecolor{golden}{rgb}{0.92, 0.7, 0}
\definecolor{midnight}{rgb}{0, 0, 0.5}
\definecolor{darkblue}{rgb}{0, 0, 0.7}

\begin{document}

\title{Fracton Matter}

\author{Andrey Gromov}
\affiliation{Department of Physics \& Condensed Matter Theory Center,
University of Maryland, College Park, MD 20742}
\email{andrey@umd.edu}
\author{Leo Radzihovsky} 
\affiliation{Department of Physics and Center for Theory of Quantum Matter,
University of Colorado, Boulder, CO 80309}
\date{\today{}}
\email{radzihov@colorado.edu}
\begin{abstract}
  We review  a burgeoning field of ``fractons'' - a class of models
  where quasi-particles are strictly immobile or display restricted
  mobility that can be understood through generalized multipolar
  symmetries and associated conservation laws. Focusing on just a
  corner of this fast-growing subject, we will demonstrate how one
  class of such theories - symmetric tensor and coupled-vector gauge
  theories surprisingly emerge from familiar elasticity of a
  two-dimensional quantum crystal. The disclination and dislocation
  crystal defects respectively map onto charges and dipoles of the
  fracton gauge theory. This fracton-elasticity duality leads to
  predictions of fractonic phases and quantum phase transitions to
  their descendants, that are duals of the commensurate crystal,
  supersolid, smectic, hexatic liquid crystals, as well as amorphous
  solids, quasi-crystals and elastic membranes.  We show how these
  dual gauge theories provide a field theoretic description of quantum
  melting transitions through a generalized Higgs mechanism. We
  demonstrate how they can be equivalently constructed as gauged
  models with global multipole symmetries.  We expect extensions of
  such gauge-elasticity dualities to generalized elasticity theories
  provide a route to discovery of new fractonic models and their
  potential experimental realizations.
\end{abstract}


\maketitle

\tableofcontents{}

\section{Introduction and motivation}
\label{intro}

Characterization and classification of phases of matter and phase
transitions between them is a central pursuit of condensed matter
physics.  The simplest and most ubiquitous organization of matter is
according to Landau's symmetry-breaking paradigm.  Such phases -
crystals, magnets, superfluids, panoply of liquid crystal phases and
many others\cite{ChaikinLubensky, deGennesProst, kleman2007soft} - are
distinguished by patterns of spontaneous symmetry breaking,
characterized by a local order parameter and a short-range entangled,
nearly product many-body wavefunction\footnote{Fermionic phases are
  more challenging with the simplest and best understood one being the
  Landau's Fermi liquid.},
%
%


Stimulated by an ever-growing class of unusual quantum materials, that
appear to lie outside of this Landau's symmetry-breaking and Fermi
liquids paradigms, much effort has been directed at exploring models,
that exhibit quantum phases with fractionalized anyonic
quasiparticles, robust spectral degeneracy sensitive only to the
topology of space, and other unusual properties common to the
so-called topological quantum liquids.\cite{SavaryBalentsQLreview}
Such exotic phenomenology is captured by conventional gauge theories,
where fractionalized quasiparticles appear at the ends of effective
field lines, free to move by growing the corresponding tensionless
string, much like charges in Maxwell's electrodynamics.
%
%

Motivated by a continued interest in topological quantum matter,
quantum glasses, and by a search of fault-tolerant quantum memory,
more recently, a new class of theoretical models has been
discovered. These feature system-size-dependent ground state
degeneracy, gapped quasiparticles with restricted mobility\footnote{No
  local operator can move an isolated excitation in one or more (or
  all) spatial directions, without creating additional excitations.},
and many other highly unusual properties.  The first, and most famous,
example is the strictly immobile excitation, dubbed ``fracton''.
Fractons and other subdimensional particles -- lineons and planeons,
restricted to move in one- and two-dimensions -- were originally
discovered in fully gapped models of commuting projector (stabilizer
codes) lattice spin
Hamiltonians\cite{chamon,bravyi,haah,cast,yoshida,haah2,fracton1,fracton2,
  bravyi11topological, haah11local, haah13thesis, haah13commuting,
  haah16algebraic}, reviewed a few years ago in
Refs.~\onlinecite{NandkishoreHermele, PCY, HET_review,
  pai19fracton}.\footnote{There are two qualitatively distinct classes
  of models with restricted mobility. As discussed in Sec. \ref{GT},
  one is with highly fragile, fine-tuned global multipolar and
  subsystem symmetries. The other are gauge-like theories with
  topological order, that are generally robust over some nonzero range
  of all local perturbations.}
%
%
%
\begin{figure}[htbp]
\hspace{-0.13in}\includegraphics*[width=0.5\textwidth]{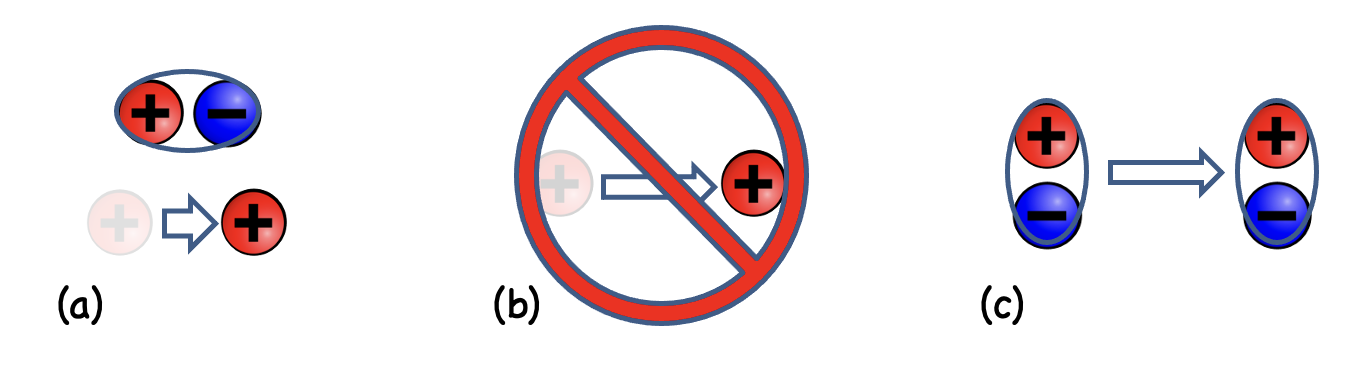}
\caption{(b) Immobility of fractonic charges enforced by dipole
  conservation illustrated in (a), with subdimensional mobility of
  dipole in (c).}
\label{immobilityFractonsFig}
\end{figure}

Such fractonic models present a challenge to our understanding of the
relationship between phases of matter and quantum field theories (QFT)
for a number of reasons \cite{QRH_fracton}. For instance, until
recently it was a piece of conventional wisdom that any gapped phase
of matter has a low-energy topological QFT (TQFT) description.
However, gapped fracton phases have a robust ground state degeneracy,
where on a spatial $d$-torus the dimension of the ground state
degeneracy grows exponentially with system\footnote{Gapped, topologically ordered fracton models can only appear in dimension $d>2$.\cite{haah_2d, aasen_2d}}
size \cite{bravyi11topological, haah11local, haah13thesis}. This is
incompatible with a TQFT description. Absence of a continuum field theory description for the Haah model can also be seen from the bifurcating nature of the renormalization group flow \cite{haah_bif}. There is, however, a description in terms of TQFT supplemented with a defect network\cite{aasen_fracton, wen_fracton}. Moreover, the number of
superselection sectors (i.e., distinct types of particle-like
excitations) also diverges. \cite{haah13commuting, haah16algebraic,
  pai19fracton} This is suggestive of an infinite number of fields
required in the continuum, with a nontrivial dependence on lattice
scale, \emph{i.e}., no obvious continuum field theory limit.

More recently, it was realized that (even if incomplete) such exotic
excitations have a natural theoretical description in the language of
higher-rank symmetric tensor gauge theories and complementary
coupled-vector gauge theories, which exhibit restricted mobility due
to an unusual set of higher (\emph{e.g.}, dipole) moment charge
conservation.\cite{subPretko, genemPretko, pretko_principle, PretkoLRdualityPRL2018,
  RHvectorPRL2020, GromovMultipole, PretkoZhaiLRdualityPRB2019} In
contrast to above models, that are based on disctete symmetries, this
class of $U(N)$ fractonic tensor gauge-theories exhibits gapless
degrees of freedom. These are related to discrete models through a
condensation of higher charge matter \cite{BB_Higgs,MaHermele}.  Rapid
recent progress in the field has established connections with numerous
other areas of physics, such as
localization\cite{abhinav,screening,circuit}, gravity\cite{mach},
holography\cite{holo}, quantum Hall systems\cite{DG_vortices,
  DMNS_APD}, and deconfined quantum criticality\cite{deconfined},
among many other theoretical
developments\cite{williamson,sagarlayer,hanlayer,parton,slagle,bowen,nonabel,balents,field,APD_phasespace,DPS_correlation,
  MSPHN_entanglement, SSC_entanglement, HZBR_EE, theta, matter,
  slagle_curved, jensen_curved, hartong_curved, XH_grav}.
\begin{figure}[htbp]
\hspace{-0.13in}\includegraphics*[width=0.5\textwidth]{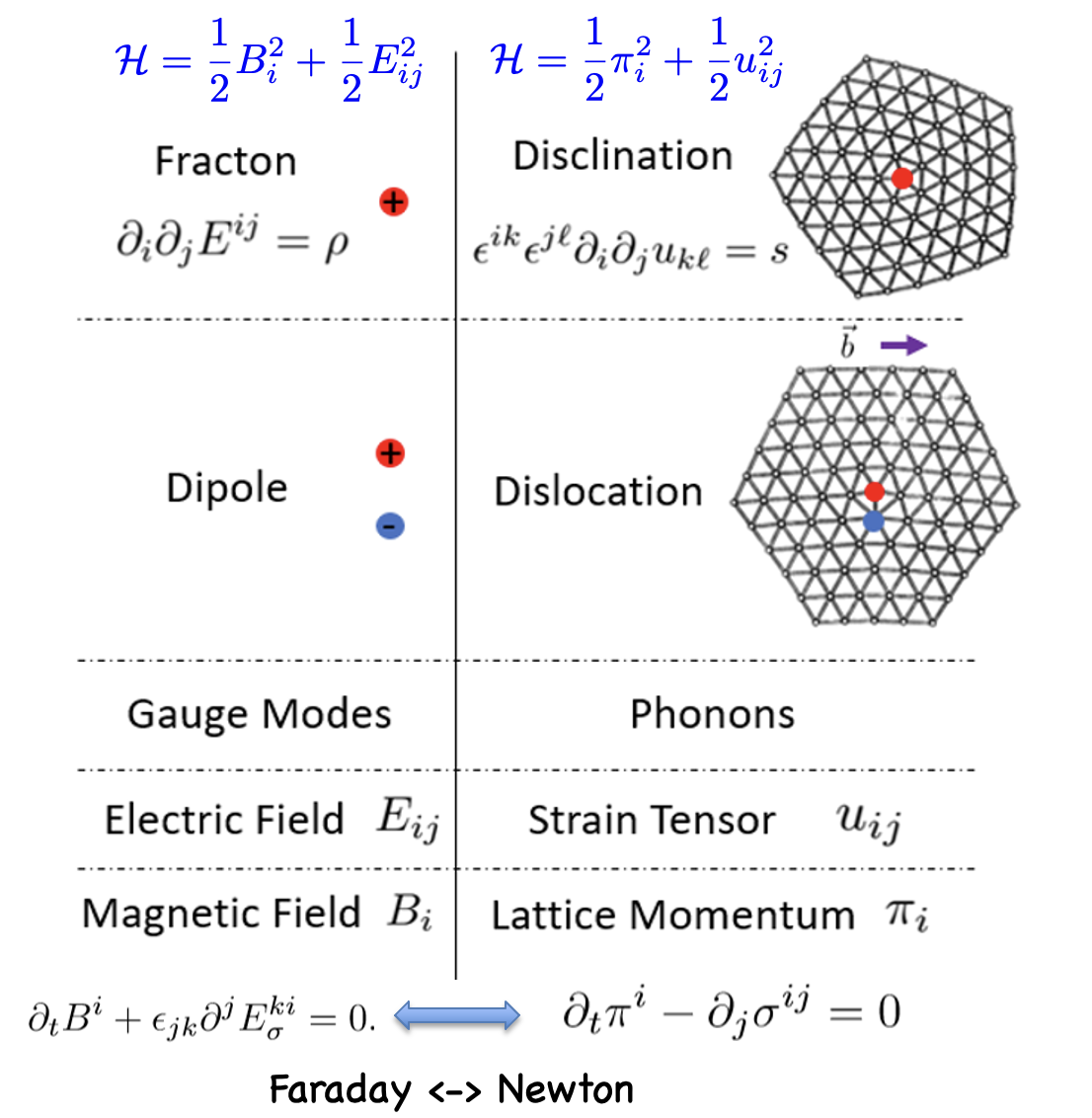}
\caption{The fracton-elasticity dictionary: topological defects,
  phonons and strains of a two-dimensional quantum crystal (right) are
  in one-to-one relation to charges, gauge fields and fields of the
  scalar-charge rank-2 tensor-gauge theory (left).}
\label{FractonElasticityDualityFig}
\end{figure}

While  exotic properties of fractons have been a subject of
intense study, concrete physical realizations have been
lacking. However, recently, it was demonstrated explicitly through
dualities\cite{Kleinert_dual} between quantum elasticity and U(1) {\em
  symmetric tensor} \cite{PretkoLRdualityPRL2018,
  PretkoLRsymmetryEnrichedPRL2018, PretkoZhaiLRdualityPRB2019,
  LRsmecticPRL2020, ZRsmecticAOP2021, KumarPotter19,
  GromovDualityPRL2019, GS_cosserat, Gromov_smectic, NGM_chiral} and
coupled {\em vector} \cite{RHvectorPRL2020, QRH_fracton} gauge
theories\footnote{Latter admit high dimensional and lattice
  generalizations.} that the fracton phenomena is realized as
topological defects in two-dimensional quantum crystals, supersolids,
and smectics.
%
%
In this Colloquium we review physics of gapless fractons, with a
particular focus on application of fractons and corresponding
fractonic gauge theories to real physical systems.  In Sec. II we
define and discuss in model-independent way unifying properties of
fractonic quasi-particles. The central component of the review appears
in Sec. III, where we present dualities between various quantum
elastic systems such crystals, supersolids and liquid crystals and
their fractonic gauge theories.  In Sec. IV we utilize these gauge
theoretic descriptions to discuss phase transitions between these
quantum phases of matter, and most interestingly a gauge-theoretic
formulation of quantum melting.  Variety of field-theoretic
constructions that give rise to gapless and gapped fracton phases are
summarized and reviewed in Sec. V. Synthesis, open questions and
future directions are relegated to the concluding Sec. VI.

\section{General perspective on restricted mobility}
\label{Def}

As we will see throughout the Colloquium, fracton excitations emerge
in a wide variety of very different physical systems. Consequently, it
is useful to decouple the phenomenon of restricted mobility of
excitations from a specific model or a physical origin that enforces
it. Thus, in this section we present a general,
realization-independent formulation of excitations with restricted
mobility, taking a symmetry-based approach.
\begin{figure}[htbp]
  \hspace{-0.13in}\includegraphics*[width=0.5\textwidth]{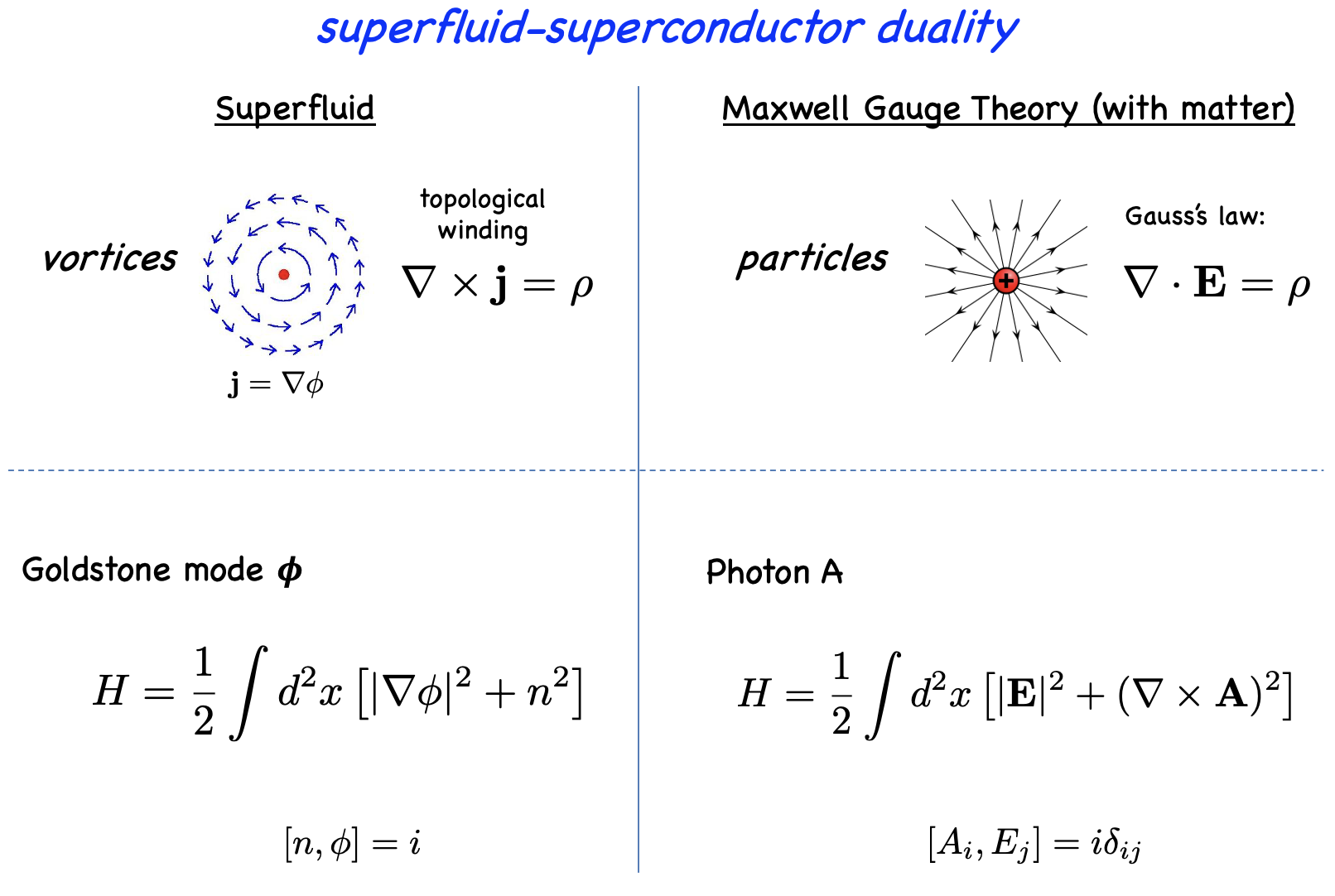}
  \caption{The boson-vortex duality in a 2+1d XY model (also known as
    superfluid-superconductor duality), whose tensor generalization is
    the fracton-disclination (gauge-elasticity) duality summarized in
    Fig.\ref{FractonElasticityDualityFig}.}
\label{bosonVortexDualityFig}
\end{figure}
In all systems we consider it will be assumed that the total number
(or charge) of fractons is conserved. The conservation can be either
exact, implemented by a $U(1)$ symmetry, or partially broken,
implemented by a symmetry breaking pattern
$U(1) \rightarrow \mathbb{Z}_p$. We start with the former.

A global $U(1)$ symmetry implies the continuity equation
\be \label{eq_continuity}
\p_0{\rho} + \p_i  J^i= 0\quad \Rightarrow \quad \p_0 Q = 0\,,
\ee
where $\rho$ is the fracton density, $J^i$ is the fracton current and
$Q(t) = \int d^d x \rho(t,\bx)$ is the total charge.  In a seminal
paper \cite{subPretko} M. Pretko showed that the mobility of charges
is restricted by further enforcing the conservation of multipole
moments $Q_{i_1 i_2\ldots i_n}$ of the charge density,
\be \label{eq_multipole}
Q_{i_1 i_2\ldots i_n} = \int d^dx \,\,x_{i_1} x_{i_2}  \ldots  x_{i_n} \rho(\bx)\,.
\ee

In order to enforce such conservation law we demand the current to be
of a special form
\be \label{eq_tensor_current}
J^i = \p_{i_1} \p_{i_2} \ldots \p_{i_{n}} J^{i_1 i_2 \ldots i_{n}i}\,,
\ee
where $J^{i_1 i_2 \ldots i_{n} i }$ is a symmetric tensor of rank
$n+1$. It describes the transport of $n$-th multipole moment
$Q_{i_1 i_2 \ldots i_{n}}$ in the direction $\hat{x}_i$. Combining
Eqs.\eqref{eq_continuity}-\eqref{eq_tensor_current} we find that all multipole moments up to the $n$-the moment are conserved (assuming that all fields decay to $0$ at infinity) 
\be\label{eq_conserv} 
\p_0Q_{i_1 i_2\ldots i_k} =0\,, \qquad k\leq n\,.
\ee
Conservation law \eqref{eq_conserv} implies restricted mobility
because generic motion of charges will change the multipole moment of
the system. The symmetry leading to these conservation laws is an
extension of spatial symmetries, dubbed multipole algebra, and has
non-trivial commutation relations with generators of rotation and
translations \cite{GromovMultipole}. Tensor and coupled vector gauge
theories discussed later emerge upon gauging this symmetry. 
%
\begin{figure}[htbp]
  \hspace{-0.13in}\includegraphics*[width=0.35\textwidth]{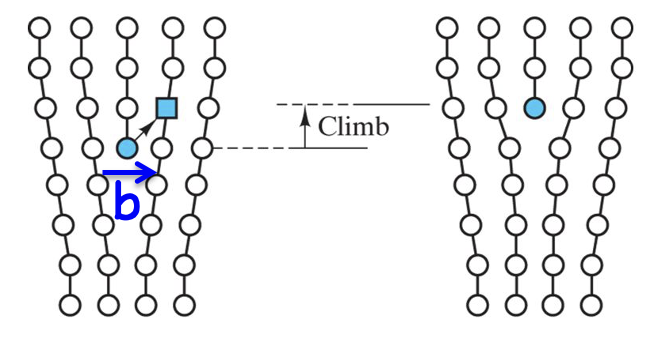}
  \hspace{-0.13in}\includegraphics*[width=0.5\textwidth]{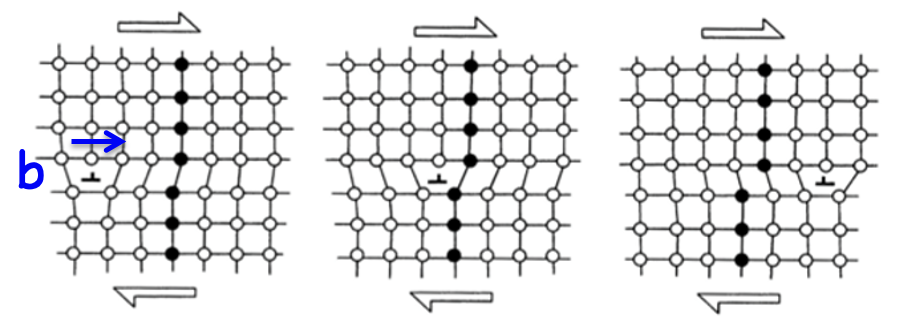}
  \caption{(a) Illustration of a dislocation climb transverse to the
    Burgers vector ${\bf b}$ (motion of a dipole along the dipole)
    forbidden by charge conservation, but made possible by a nonzero
    density of vacancies and interstitials. (b) Allowed dislocation
    lineon glide motion along ${\bf b}$ (transverse to the dipole).}
  \label{climbGlideFig}
\end{figure}

To make this generic formulation more concrete we consider a theory
with a conserved dipole moment, $Q_i$, and examine its two fracton
excitations of equal and opposite charge located at $\bx_1$ and
$\bx_2$. Let $T_i({\bf a}_i)$ be an operator that translates
$i$-th fracton by vector ${\bf a}_i$. Then dipole conservation implies
\be \label{eq_dipole1}
\langle \bx_1, \bx_2 | T_1({\bf a}_1) T_2({\bf
  a}_2)|\bx_1, \bx_2 \rangle \propto
\delta^{(d)}({\bf a}_1 - {\bf a}_2)\,. 
 \ee
The same conclusion can also be reached directly from the constraint
on the current $J^i = \p_j J^{ji}$. For simplicity focussing on 2d, we
consider a wide strip, extended in $\hat{x}_2$ direction, and assume
that the system is in a homogeneous steady state. The total charge
current flowing through a cross section of a strip in $\hat{x}_1$
direction is
\be
J^i_{tot} = \int_{-\infty}^{\infty} dx_2 (\p_1 J^{1i} + \p_2 J^{2i}) = \p_1 \int_{-\infty}^{\infty} dx_2 J^{1i} = 0\,,
\ee
 where in the last step we used
homogeneity. Indeed, if the charges can only move around in the form
of bound dipoles, the total current through any cross section in a
homogeneous state will always be $0$.

Complementarily, we can define a microscopic dipole current
${\tilde J}^{ij} = x^i J^j$, with the dipole density $\rho^i$ satisfying
\be\label{eq_dipolecontinuity}
\p_0\rho^i + \p_j {\tilde J}^{ij}  = J^i\,.
\ee
This demonstrates that motion of monopoles ($J^i \neq 0$) generates
dipoles, thereby violating their continuity equation, and equivalently
dipole conservation demands a vanishing of monopole current,
$J^i = 0$, i.e., immobility of fractonic
monopoles \cite{RHvectorPRL2020, LRsmecticPRL2020}.

The conservation laws discussed above do not cover all variety of
mobility-constrained systems. There are two important generalizations
one has to consider. First, Eq.\eqref{eq_tensor_current} assumes that
the tensor current transforms in a representation of $SO(d)$. Since
most systems supporting fractons are initially formulated on a
lattice, there is no \emph{a priori} reason for $SO(d)$ symmetry to be
relevant. Indeed, the current may transform in representations of a
point group symmetry. This is crucial to fit lattice models with
discrete symmetries, like the X-cube, Chamon and Haah codes into this
framework. Second, the charge density $\rho$ itself can be assumed to
be a tensor transforming in some representation of either $SO(d)$ or
its point subgroup. In this case the relevant multipole moments are
the elementary excitations and cannot be divided further into smaller
moments or point charges. Finally, when the charge conservation is
broken down to $\mathbb{Z}_p$, the multipole conservation still
imposes modulo-$p$ constraints on the system. However, a general theory
of such systems is not yet developed.

\section{Fractons as crystalline defects}
\label{Defects}
In this Section we describe an intriguing connection of fractonic
dynamics to the restricted mobility of positional (dislocations) and
orientational (disclinations) topological defects in a familiar
quantum two-dimensional crystal \cite{PretkoLRdualityPRL2018,
  PretkoLRsymmetryEnrichedPRL2018, PretkoZhaiLRdualityPRB2019,
  GromovDualityPRL2019, RHvectorPRL2020}. Thus, a quantum crystal
provides the only physical realization of a fractonic system known
to date.  Building on this we will further show how the corresponding
coupled U(1) vector (and the equivalent symmetric tensor-) gauge
theory provides a theory of quantum melting of a crystal and discuss
intermediate liquid crystal phases. This Section will introduce the
main ideas from the theory of particles with restricted mobility as
well as tensor and multipole gauge theories.

\subsection{Particle-vortex duality}
\la{PVD}
As a warm up for fractonic duality of crystals and liquid crystals we
briefly review the standard 2+1d particle-vortex, or
equivalently\footnote{It is also known as a duality between a
  superfluid and a superconductor.}, XY-to-Abelian-Higgs model
duality\cite{Peskin, DHduality,FisherLee}.
\begin{figure}[htbp]
  \hspace{0in}\includegraphics*[width=0.3\textwidth]{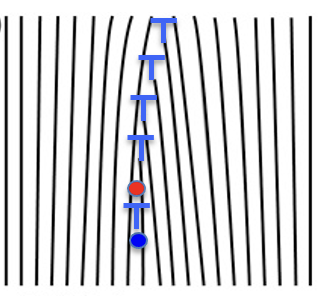}
  \caption{Instantiation of an immobile fracton as a disclination
    defect of a 2+1d crystal. Viewing a disclination as an end point
    of a ray of dislocation dipoles, its hop by a lattice constant
    (from red to blue dot) corresponds to an addition of a dislocation
    dipole. Since later corresponds to an addition of a half-ray of
    atoms, such highly nonlocal operator is not allowed in a local
    Hamiltonian (and not just by a global symmetry), thereby strictly
    forbidding disclination motion.}
\label{immobileDisclinationFig}
\end{figure}
The low energy effective Lagrangian describing 2+1d bosons is given
by\footnote{The XY model Lagrangian and its duality can more carefully
  be treated on a lattice, giving equivalent results. Thus, our
  streamlined continuum approach is fundamentally justified by lattice
  regularization. }
\be \la{eq_eft_sf}
\mathcal L = \frac{K}{2} (\partial_\mu\phi)^2\,,
\ee
where $\phi$ is the superfluid phase, and we used the units in which
the speed of sound is $1$. In the presence of vortices, the boson
current $\partial_\mu\phi$ can be decomposed into a smooth and
singular (vortex) parts,
$\partial_\mu\phi = \partial_\mu\phi_s + v_\mu$. The circulation
of the latter,
\be
\epsilon^{\mu\nu\rho}\p_\nu v_\rho \equiv (\p\times v)^\mu = j_v^\mu,
\label{jvortex}
\ee
gives the vortex 3-current $j^v_\mu = (\rho^v, j_i^v)$.

Introducing a Hubbard-Stratonovich field $J_\mu$, transforms the
Lagrangian \eqref{eq_eft_sf} into
\be
\mathcal L = \frac{K^{-1}}{2} J_\mu J^\mu + (\p_\mu\phi_s + v_\mu) J^\mu\,. 
\ee
Integrating out the smooth component of $\phi$ enforces the boson
continuity equation
\be
\p_\mu J^\mu=0\,,
\ee
solved in terms of a vector potential
$J^\mu = \epsilon^{\mu\nu\rho}\p_\nu A_\rho=(\partial\times A)^\mu$,
with the physical current invariant under a gauge transformation
\be\la{eq_u1}
\delta A_\mu = \p_\mu \chi\,.
\ee 
The Lagrangian \eqref{eq_eft_sf} then takes a form of a U(1) gauge
theory
\begin{eqnarray}
\tilde{\mathcal L} &=&
\frac{K^{-1}}{2}(\pt\times A)^2 + j_v^\mu A_\mu\,,\nonumber\\
&=& \frac{K^{-1}}{2}(E^2 - B^2) + j_v^\mu A_\mu\,,
\label{dualLxy}
\end{eqnarray}
where the electric field $E_i = \p_0 A_i - \p_i A_0$ and the magnetic
field $B = \epsilon^{ij} \p_i A_j$ respectively describe the spatial
components of bosonic current and number densities, and under the
duality the vortex 3-current $j_v^\mu$ enters Lagrangian as the dual
matter.

In an equivalent Hamiltonian description, illustrated in
Fig.\ref{bosonVortexDualityFig}, bosons are described by
\be
\hat\cH = \oh|\grad\hat\phi|^2 + \oh\hat
\rho^2,
\ee
in terms of canonically conjugate $[\hat\rho,\hat\phi] =
i\delta^{(2)}(\xv)$, and vortex singularities, $\grad\times\grad\phi =
\rho_v$. On the dual side, rotating the boson current
$\grad\phi$ lines by
$\pi/2$ transforms them into electric field lines
$\E$ and boson density into dual flux density
$\B$, with $[\hat\A,\hat\E] =
i\delta^{(2)}(\xv)$.  This then transforms the bosonic Hamiltonian into
the dual Maxwell theory
\be
\tilde{\hat\cH} = \oh|\hat\E|^2 +
\oh(\grad\times\hat\A)^2 -
\jv_v\cdot\A\,.
\ee
It is more formally obtained from the dual Lagrangian \rf{dualLxy} by
introducing an independent electric field $\bf E$ as the
Hubbard-Stratonovich field to decouple the spatial (electric field
energy) part of the Maxwell term.  The
Integration over the time-component $A_0$ then gives the Gauss's law,
\be
\grad\cdot\E = \rho_v\,,
\label{Gauss}
\ee
that generates \eqref{eq_u1} and is the dual counter-part of the
circulation constraint, \rf{jvortex}.


We thus recover the celebrated Dasgupta-Halperin duality\cite{Peskin,
  DHduality, FisherLee}, where a bosonic liquid is dual to vortex
matter, described by a complex scalar field, $\Psi$, minimally coupled
to a $U(1)$ gauge field $A_\mu$,
\bea
\mathcal L &=& i\Psi^\star \left(\p_0 - i A_0\right)\Psi - \frac{1}{2m}
|\left(\p_i - i A_i\right) \Psi|^2 + V(|\Psi|) \nonumber\\
&&+ \frac{K^{-1}}{2}(\partial\times A)^2\,.
\eea
It is also known as the Abelian-Higgs model of a dual superconductor,
with $V$ a generic $U(1)$ symmetric potential.  In this duality, a
superfluid and Mott insulating phases of bosons thus respectively
correspond to a dual-normal (dual-non-superconducting, $\Psi=0$) state and
a dual-superconducting vortex condensate ($\Psi \neq 0$ dual-Higgs) phase.

\subsection{Fracton-disclination duality: commensurate crystal}

\subsubsection{Tensor gauge theory duality}
\la{sec:ElasticDuality}

The duality of an elastic medium is an eligant, technically
straightforward tensor generalization\cite{Kleinert_dual,
  kleinert1,kleinert2,zaanen} of the above XY-duality, where the
phonons $u_i$, dislocations and disclinations are respective vector
counter-parts of the superfluid phase $\phi$ and vortices. Since 2d
elastic medium is described by phonon Goldstone modes that are spatial
vectors, it is not surprising that their dual is captured by a tensor
(rather than vector) gauge field $A_{ij}$.\footnote{See however,
  Sec.\ref{coupledVector} for the coupled U(1) vector gauge theory
  formulation of elasticity dual, at low energies equivalent to the
  tensor gauge theory presented in this section.}

\begin{figure}[htbp]
  \hspace{-0.13in}\includegraphics*[width=0.5\textwidth]{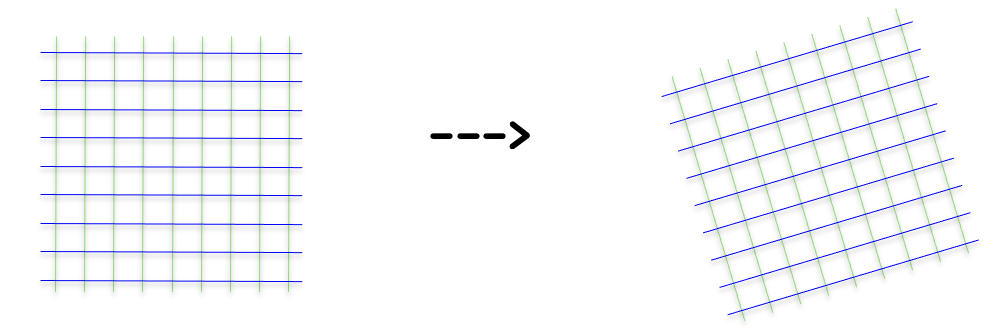}
  \caption{Target-space rotational symmetry of a crystal, encoded in a
    symmetric strain tensor $u_{ij}$ appearing in its elastic energy.}
\label{targetRotationXtalFig}
\end{figure}

To this end \cite{PretkoLRdualityPRL2018,
  PretkoLRsymmetryEnrichedPRL2018, PretkoZhaiLRdualityPRB2019}, we
begin with a low-energy elastic description of a $2+1$ dimensional
quantum crystal, captured by a harmonic Lagrangian,
\be
\label{Lcrystal}
\mathcal L = \oh(\pt_0 u_i)^2 - \oh C^{ijkl} u_{ij} u_{kl}\,,
\ee
where $u_{ij} = \frac{1}{2} \left(\p_i u_j + \p_j u_i \right)$ is the
linearized symmetric strain tensor (encoding target-space rotational
invariance in Fig. \ref{targetRotationXtalFig}), rank-$4$ tensor
$C^{ijkl}$ encodes elastic moduli and underlying crystal symmetry, and
we specialized to unit mass density \cite{ChaikinLubensky}. To include
topological defects, we express the displacement in terms of smooth
phonon and singular defects components
$u_i = u_i^{\rm s} + u_i^{\rm d}$, with latter accounting for
disclination density $\rho$,
\begin{equation}
  \epsilon^{ik}\epsilon^{j\ell}\partial_i\partial_ju_{k\ell} =
  \rho\, ,
\label{disc}
\end{equation}
and implicitly for dislocations as they are dipole bound states of two
disclinations \cite{ChaikinLubensky,landau,SeungNelson1988}.  Introducing the
Hubbard-Stratonovich momentum, $\pi^i$ and symmetric stress
$\sigma^{ij}$ fields and integrating over the phonons $u_i^{\rm s}$,
enforces momentum continuity $\p_0\pi^i + \p_j\sigma^{ij} = 0$. To
make contact with electromagnetism, it is convenient to introduce dual
vector magnetic $B_i=\epsilon^{ij}\pi_j$ and {\em symmetric} tensor
electric $E^{ij}_{\sigma}=\epsilon^{ik} \epsilon^{jl}\sigma_{kl}$
fields
%
%
in terms of which the momentum continuity equation takes the form of a generalized
Faraday law,
\begin{equation}
\p_0 B^i + \epsilon_{jk}\partial^j E_\sigma^{ki} = 0
\label{Faraday}
\end{equation}
of the scalar-charge tensor gauge theory \cite{subPretko}. As in a
conventional electromagnetism, latter can be solved in terms of gauge
fields, here symmetric tensor, $A_{ij}$ and scalar $A_0$,
%
%
%
\be \label{eq_EB}
B_i = \epsilon^{kl} \p_k A_{li}\,, \qquad E^{ij}_{\sigma} = -\p_0 A^{ij} + \p^i \p^jA_0\,,
\ee
invariant under the \emph{gauge} transformations
\be \label{eq_gauge_tr}
\delta A_{ij} = \p_i \p_j \chi\,, \quad \delta A_0 = \p_0{\chi}\,.
\ee
The elasticity Lagrangian \eqref{Lcrystal} then takes on a
Maxwell-like form
\be\la{eq_gMaxwell} \mathcal L = \oh\tilde{C}_{ijkl} E^{ij}
E^{kl} - \oh B^2 + \rho A_0 - J^{ij} A_{ij}\,,
\ee
where
$\tilde{C}^{-1}_{ijkl} =
C^{-1}_{mnpq}\epsilon_i{}^m\epsilon_j{}^n\epsilon_k{}^p\epsilon_l{}^q$ and $E_{ij} = \tilde{C}^{-1}_{ijkl}E^{kl}_\sigma$. 
The dual gauge fields are sourced by crystalline topological defects
through the last two terms, \eqref{eq_gMaxwell} with dislocation
current given by
\begin{equation}
J^{ij} =
\epsilon^{ik}\epsilon^{j\ell}(\partial_0\partial_k
- \partial_k\partial_0)u_\ell\, . 
\end{equation}
Gauge invariance \eqref{eq_gauge_tr} enforces that the dual charge and
currents densities $\rho$ and $J^{ij}$ satisfy a continuity equation
\be\label{eq_elast}
\p_0{\rho} + \p_i \p_j J^{ij} = 0\,,
\ee
with a current indeed constrained in a manner discussed in Section
\ref{Def}, as required on general grounds.  Notably, the immobility of
fractonic disclination charges $\rho$ is reflected in the absence
(vanishing) of fracton current. Gauss's law generating the gauge
transformations \eqref{eq_gauge_tr} can be read out from
\eqref{eq_EB},\eqref{eq_gMaxwell} and appears as a constraint after
integration over the scalar potential $A_0$ and takes the form
\be\la{eq_TSCT_Gauss}
\p_i \p_j  E^{ij} = \rho\,.
\ee
It is the counterpart of the topological disclination condition
\rf{disc}, as illustrated in Fig.\ref{FractonElasticityDualityFig}.

For an isotropic (e.g., triangular) crystal, the elastic tensor
$C_{ijkl}$ reduces to two independent Lam\'e moduli $\mu$ and
$\lambda$ and dual Maxwell-like theory \eqref{eq_gMaxwell} exhibits
two propagating gapless degrees of freedom corresponding to transverse
and longitudinal phonons. This demonstrates that a quantum
crystal is dual to the so-called \emph{traceless scalar charge} gauge
theory
\cite{PretkoLRdualityPRL2018,PretkoLRsymmetryEnrichedPRL2018,PretkoZhaiLRdualityPRB2019,GromovDualityPRL2019},
and gives the only known physical realization of fractonic matter.

%
%

We note that a conservation of the number of occupied lattice sites,
i.e., absence of vacancy and interstitial point defects,
characteristic of a ``commensurate'' crystal implies that $J^{ij}$ is
\emph{traceless}, further constraining the motion of
defects\cite{PretkoLRsymmetryEnrichedPRL2018,PretkoZhaiLRdualityPRB2019,KumarPotter19}. Equation \eqref{eq_elast}
then implies that total dual charge, dipole moment and trace of the
quadrupole moment are conserved
\be
\label{conserv}
\p_0Q = 0\,, \qquad \p_0Q_i = 0\,, \qquad
\p_0\tr\left(Q_{ij}\right) = 0\,,
\ee
Identification of charges, dipoles and quadrupoles of this fracton
phase F$_{U(1)}$ with disclinations, dislocations (with Burgers vector
perpendicular to the dipole moment) and
vacancies/interstitials\cite{PretkoZhaiLRdualityPRB2019,
  PretkoLRsymmetryEnrichedPRL2018, KumarPotter19,GromovDualityPRL2019}
gives an elegant gauge theory formulation of their quantum dynamics in
a ``commensurate'' crystal. Namely, the fracton charges encode the
immobility of disclinations, the duality
prediction\cite{PretkoLRdualityPRL2018} that can also be understood
directly in terms of crystal degrees of freedom, as illustrated in
Fig. \ref{immobileDisclinationFig}. The planeon dipoles, with motion
constrained (by the $U(1)$ conservation of the trace of quadrupoles in
\rf{conserv}) transverse to the dipole moment, correspond to the
well-known \emph{glide-only constraint}, that, in the absence of
vacancies and interstitials prevents dislocations from \emph{climbing}
perpendicular to their Burgers
vector\cite{PretkoLRsymmetryEnrichedPRL2018, PretkoZhaiLRdualityPRB2019,KumarPotter19,GromovDualityPRL2019},
as illustrated in Figs.\ref{climbGlideFig}(a) and
\ref{dipoleMobilityFullFig}.  In Sec. \ref{supersolid}, we will
discuss in more detail the relaxation of this $U(1)$ symmetry-enriched
constraint, that leads to a qualitatively distinct fractonic state F,
corresponding to an {\em incommensurate} crystal, i.e., a supersolid.

\subsubsection{Disclination field theory}

With crystalline defects appearing as dual matter that sources tensor
gauge fields, the formulation extends to a concrete field theoretic
representation of fractonic matter, $\rho$ and $J^{ij}$. To this end,
we describe disclinations by a complex scalar field $\Phi$, treating
them as bosonic excitations. The Lagrangian
$\cL = \cL_s[A_{ij},\Phi] + \cL_{Max}[A_{ij}]$ is a sum of the dual
disclination matter sector, $\cL_s$ and the Maxwell-like sector,
$\cL_{Max}$, derived in the previous subsection, capturing the phonon
degrees of freedom. The dual matter Lagrangian, $\cL_s$ takes the form
\cite{pretko_principle}
\be\la{eq_discl_L}
\mathcal L_s = i\Phi^*\left(\p_0 - i A_0\right)\Phi + g \left(|D_1(\Phi)|^2 + |D_2(\Phi)|^2\right) + V(\Phi)\,,
\ee
where $D_I(\Phi)$ are differential operators, \emph{bilinear} in
$\Phi$, given by
\be\la{eq_diff}
D_I(\Phi) = \Pi_I^{ij} \left( \p_i \Phi \p_j \Phi - \Phi \p_i \p_j - i A_{ij} \Phi \right)\,, \quad \vec{\Pi} = (\sigma_1,\sigma_3)\,,
\ee
where $\sigma_1, \sigma_3$ are the Pauli matrices.  In the absence of
gauge fields, the Lagrangian \eqref{eq_discl_L} exhibits a conservation
law \eqref{eq_elast} due to the presence of an unusual ``global''
symmetry, 
\be\la{eq_global}
\Phi^\prime = e^{i\chi_g(\xv)} \Phi\,, \qquad \chi_g(\xv) = \alpha
+ \beta_i x^i + \oh\gamma |x|^2\,, 
\ee
characterized by
phase $\chi_g(\xv)$, which by the virtue of Noether theorem leads to
the conservation of $U(1)$ charge, dipole and trace of the quadrupole
moments.  In the presence of tensor gauge fields, transforming
according to \eqref{eq_gauge_tr}, the Lagrangian \eqref{eq_discl_L} is
also invariant under a general gauge transformation
$\Phi\rightarrow e^{i\chi(\xv)}\Phi$, with arbitrary $\chi(\xv)$.  As
in conventional gauge theories, here the disclination density and
dislocation current are given as variational derivatives
\be
\rho = \frac{\delta S}{\delta A_0}\,, \qquad J^{ij} = \frac{\delta S}{\delta A_{ij}}\,.
\ee
It follows from \eqref{eq_diff} that the current $J^{ij}$ is indeed
traceless as expected\footnote{More precisely, $J^{ij}$ transforms in
  spin-$2$ irreducible representation of $SO(2)$.}.

\subsubsection{Dislocation field theory}

The disclination field theory, $\cL_s$ is \emph{quartic} in the fields
and thus does not admit a weak-coupling quadratic representation,
reflecting its strongly interacting UV degrees of freedom. Indeed, in
real crystals disclinations appear in tightly bound lattice-scale
dislocation dipoles. To construct their field theory, $\cL_d$, we
observe that a dislocation with a Burgers vector $\bv$ (an elementary
lattice vector) is created by an operator $\psi^\dagger_{\bf d}$,
labeled by dislocation dipole $\bf d$ (corresponding to $\rho_i$
density of Sec. \ref{Def} and $\pv$ in the following sections), with the
corresponding coherent state field,
\be\la{eq_dipole} \psi^\star_{\bf d}(\xv) =
\Phi\left(\xv-\frac{\dv}{2}\right)
\Phi^\star\left(\xv+\frac{\dv}{2}\right)\,,
\ee
and is bi-local in the disclination fields, with $\bv = {\bf
  d}\times\zh$. Under the global symmetry \eqref{eq_global}, in the
$d\rightarrow 0$ limit, the dipole operator transforms as 
\be\la{eq_dipole_sym}
\psi_{\bf d}(\xv) \rightarrow e^{i \beta_i d^i  + i \gamma d_i x^i} \psi_{\bf d}(\xv)\,,
\ee
where the first factor is a global phase enforcing $U(1)$ symmetry,
while the second factor enforces the glide-only constraint.

In constrast to a strongly interacting disclination Lagrangian
\eqref{eq_discl_L} (that has no noninteracting limit), mobile
dislocations admit a weakly interacting Lagrangian. Its form is
constrained by the generalized global (subsystem) symmetry
\eqref{eq_dipole_sym} as well as the gauge symmetry
\eqref{eq_gauge_tr} and is uniquely given by \cite{KumarPotter19,PretkoZhaiLRdualityPRB2019},
\begin{multline}
\mathcal L_d = \sum_{\bf d}i\psi_{\bf d}^\star \Big(\p_0 - iA_0
\Big)\psi_{\bf d}  
\\
- \frac{P^{ij}_\perp}{2m}\left(\p_i + i d_k A_{ik}\right)\psi_{\bf d}^\star
\left(\p_j - i d_l A_{jl}\right)\psi_{\bf d} + V(|\psi_{\bf d}|^2)\,,
\label{tensorGLcrystalDual}
\end{multline}
%
%
%
where
$P^{ij}_\perp = \left(\delta^{ij} - \frac{d^i d^j}{|d|^2}\right)$ is
the projector onto axis perpendicular to the dipole moment $\dv$
(\emph{i.e.}, along the Burgers vector, $\bv = \zh\times\dv$),
enforcing the glide-only constraint, dictated by
\eqref{eq_dipole_sym}. The dipole moment enters as the ``charge'' of
the field $\psi_d$ under the tensor gauge field $A_{ij}$.  The
Lagrangian \eqref{tensorGLcrystalDual} is also invariant under
discrete transformations that permute primitive lattice vectors,
corresponding to the point group symmetry of the crystall.

\subsubsection{Coupled vector gauge theory duality}
\label{coupledVector}

A complementary, more convenient and transparent formulation of
crystal's dual gauge theory is that in terms of coupled U(1) {\em
  vector} gauge fields, introduced and detailed in
Ref.~\onlinecite{RHvectorPRL2020}. The idea is in fact quite simple,
and is based on the observation that elasticity, formulated in terms
of the (unsymmetrized) strain $\partial_i u_k$ has a form of two
flavored $u_x$ and $u_y$ XY-models. Of course independent XY-models
would dualize to conventional non-fractonic U(1) vector gauge
theories. Thus, it is the nontrivial ``flavor-space'' phonon coupling
(symmetrization of $\partial_i u_k$ in the conventional formulation)
that is responsible for the appearance of
fractons. \cite{RHvectorPRL2020, QRH_fracton}

To get to an equivalent flavored {\em vector} gauge theory
description, we reformulate the conventional elastic theory
\rf{Lcrystal} in terms of ``minimally''-coupled quantum XY-models,
introducing the orientational bond-angle field, $\theta$ and its
canonically conjugate angular momentum density $L$. The Lagrangian
density (for simplicity taking elastic constants tensor $C_{ijkl}$ to
be characterized by a single elastic constant $C$) is given by
\begin{eqnarray}
\cL &=&  \pi_k\p_0 u_k + L\partial_0\theta - \oh\pi_k^2 - \oh C(\pt_i  u_k -   \theta\epsilon_{ik})^2
\nonumber \\ 
&-& \oh  L^2 - \oh K(\grad\theta)^2. \label{Lutheta}
\end{eqnarray}
Coupling of the unsymmetrized strain $\pt_i u_k$ to
$\theta\epsilon_{ik}$, ``Higgses'' its anti-symmetric part below a
scale set by $C$, reducing $\cL$ to standard crystal elasticity in
terms of the symmetrized strain $u_{i k}$ \rf{Lcrystal}, which is the
starting point of Ref.~\onlinecite{PretkoLRdualityPRL2018}.  This
reformulates 2d elasticity in terms of two translational XY models for
two phonons, $u_k$, coupled by the orientational XY model for the
orientational bond field, $\theta$.  To dualize $\cL$, we decouple the
elastic and orientational energies in (\ref{Lutheta}) via
Hubbard-Stratonovich vector fields, stress $\vsigma_{k}$ (with flavor
index $k$ inherited from $\grad u_k$) and torque $\vtau$.
We then introduce disclinations
$\grad\times\grad\theta^s = \frac{2\pi
  s}{n}\delta^2(\rv)\equiv\rho(\rv)$ with charge $2\pi s/n$
($s = \pm \mathbb{Z}$ and $n = 6$ for hexagonal crystal) and their
dipoles, dislocations,
$\grad\times\grad u^s_k = b_k\delta^2(\rv)\equiv b_k(\rv)$ with
Burgers charge $\bv$, and integrate over the single-valued elastic
components of $\theta$ and $u_k$. This enforces the conservation of
linear and angular momenta,
$\p_0\pi_k - \grad\cdot\vsigma_{k} = 0$ and
$\p_0 L - \grad\cdot\vtau =\eps_{ij}\sigma_{ij}\equiv\sigma_a$.

Expressing this linear momentum constraint in terms of dual magnetic
and electric fields, $\pi_k = \epsilon_{kj}B_j$,
$\sigma_{ik} = -\epsilon_{ij}\epsilon_{k\ell}E_{j\ell}$, gives
$k$-flavored Faraday equations,
$\p_0 B_k + \grad\times {\bf E}_{k} = 0$, solved by $k$-flavored
vector ${\bf A}_{k}$ and scalar $A_{0k}$ gauge potentials,
$B_k = \grad\times{\bf A}_{k}$,
${\bf E}_{k} = -\p_0 {\bf A}_{k} - \grad A_{0k}$. We emphasize
that, in contrast to the {\em symmetric tensor}
approach\cite{PretkoLRdualityPRL2018,PretkoLRsymmetryEnrichedPRL2018},
here, the $k=(x,y)$-flavored {\em vector} gauge field ${\bf A}_{k}$
has components $A_{ik}$ that form an unsymmetrized tensor field.

The conservation of angular momentum can now be solved with another
set of vector ${\bf a}$ and scalar $a_0$ gauge fields, 
\begin{eqnarray}
L &=& \grad\times{\bf a} + A_a,\ \
\tau_k = \epsilon_{kj}(\pt_0 a_j + \pt_j a_0 - A_{0j}),\qquad
\label{solveL}
\end{eqnarray} 
leading to the dual Lagrangian density,
\begin{eqnarray}
\label{LdualCr}
\tilde\cL_{\text cr} &=& \oh C^{-1}(\p_0 {\bf A}_k + \grad A_{0k})^2
- \oh(\grad\times {\bf A}_{k})^2\\
&+& \oh K^{-1} (\p_0 a_k + \partial_k a_0 - A_{0k})^2
- \oh(\grad\times {\bf a} + A_a)^2\nonumber \\
&+& {\bf A}_{k}\cdot{\bf J}_{k} -A_{0k}p_k
+{\bf a}\cdot{\bf j} - a_0\rho \,,\nonumber
\end{eqnarray}
where the dipole charge $p_k$ is given by the dislocation density
$p_k = (\zh\times{\bf b})_k$, the fracton charge $\rho$ is the
disclination density, and the corresponding currents are
given by ${\bf J}_{k} = \epsilon_{lk}\zh\times (\p_0\grad u_l - \grad\p_0 u_l)$ and 
${\bf j} =  \zh\times(\p_0\grad\theta - \grad\p_0\theta)$.

The corresponding Hamiltonian density 
\begin{eqnarray}
\label{eqn:mainL}
  \tilde{\cH}_{\text cr} &=&
  \oh C |{\bf  E}_k|^2 + \oh(\grad\times {\bf  A}_{k})^2
  +\oh K |{\bf  e}|^2\nonumber\\
  &&+ \oh (\grad\times{\bf  a} +  A_a)^2 
  -{\bf  A}_{k}\cdot{\bf J}_{k} - {\bf a}\cdot{\bf j}\; ,\;\;\;
\end{eqnarray}
involves three U(1) vector gauge fields with electric fields
${\bf E}_k$ (flavors $k = x,y$) and ${\bf e}$, and corresponding
canonically conjugate vector potentials ${\bf A}_k$ and ${\bf a}$,
with former gauging the latter through a 2-form
$A_a = \epsilon_{ik} A_{ik}$ minimal coupling, $(da - A)^2$.  This
translational and orientational gauge fields coupling encodes a
semi-direct product of spatial translations and rotations, constrained
by the generalized gauge
invariance \cite{RHvectorPRL2020}.

The Hamiltonian is supplemented by Gauss's laws,
\begin{eqnarray}
\grad\cdot{\bf E}_k = p_k -  e_k\,,\label{Coulomb_p}\qquad
\grad\cdot{\bf  e}= \rho\,. \label{Coulomb_rho}
\end{eqnarray}
Crucially, the components of the electric field $ e_k$ (that would be
generated by the motion of charges $\rho$) appear as an additional
dipole charge in the dipole Gauss's law for $\E_k$,\rf{Coulomb_p}, and
its conservation thus encodes the fractonic immobility of disclination
charges, $\rho$.  Equivalently, we note that the continuity equation
for dipole densities $p_k$,
\begin{equation}
 \partial_0 p_k + \grad \cdot {\bf J}_k={\bf j}\;,\ \ \ \
 \partial_0 \rho + \grad \cdot {\bf j} = 0,
\label{ContinuityCr}
\end{equation}
%
%
is violated by the charge current $\jv$, which thus must vanish in the
absence of dipole charges, i.e,. $p_k=0\rightarrow j_k = 0$, leading
to immobility of fractonic charges, enforced by gauge-invariance.

Although above elasticity-gauge duality only works in 2+1d dimensions,
the generalization of the gauge dual to $d$ dimensions is
straightforward (though does not correspond to any physical
elasticity) and consists of $d+1$ U(1) gauge fields obeying the same
Gauss's laws but with $k = 1,\dots, d$.  The main difference in the
Hamiltonian is that $(\grad\times{\bf a} + A_a)^2$ is replaced by a
sum of the $d(d-1)/2$ terms of the 2-form minimal coupling
$(\partial_i a_j - \partial_j a_i + A_{i j} - A_{j i})^2 = (da -
A)^2$.  At low energies this coupled $U(1)$ vector gauge theory
reduces to the $d$-dimensional scalar-charge tensor gauge
theory \cite{RHvectorPRL2020}.



\subsection{Fracton-disclination duality: smectic}
\label{smecticDual}
A smectic state of matter, characterized by a uniaxial {\em
  spontaneous} breaking of rotational and translational symmetries is
ubiquitous in classical liquid crystals of highly anisotropic
molecules (e.g., classic 5CB).\cite{deGennesProst} Its quantum
realizations range from ``striped'' states of a two-dimensional
electron gas at half-filled high Landau
levels\cite{EisensteinQSm,CsathyARCMP,Fogler,Moessner2,FradkinKivelsonQHsm,FisherMacdonald,LR_Dorsey, barci2002theory},
and ``striped'' spin and charge states of weakly doped correlated
quantum magnets\cite{TranquadaStripes,KivelsonStripes} to the putative
Fulde-Ferrell-Larkin-Ovchinnikov (FFLO) paired superfluids\cite{FF,LO} in
imbalanced degenerate atomic gases\cite{LR_VishwanathPRL,LRpra},
ferromagnetic transition in one-dimensional spin-orbit-coupled
metals\cite{KoziiPRB17}, spin-orbit coupled Bose
condensates,\cite{LR_ChoiPRL,HuiZhai} as well as helical states of
bosons or spins on a frustrated lattice.\cite{helicalBosonsSMR22} We
thus next review a smectic dual gauge theory representation and the
associated quantum melting transitions from surrounding crystal and
nematic phases.

A simplest description of a $2+1$d quantum
smectic\cite{LRsmecticPRL2020,ZRsmecticAOP2021} is in terms of a
single phonon Goldstone mode, with a Lagrangian density,
\begin{eqnarray}
  \mathcal{L}_{\text{sm}}=\frac{1}{2}(\p_0 u)^2
  -\frac{\kappa}{2}(\partial_y u)^2 -
  \frac{K}{2}(\partial^2_x u)^2,
\label{LsmLifshitz}
\end{eqnarray}
a close cousin of the quantum Lifshitz model
\cite{mLifshitz,VishwanathBalentsSenthil2004, Z3LRsineGordonPRB,
  FradkinHuse2004, ArdonneFendleyFradkin2004, LakeDBHM, KS_HMW_dipole,
  ShuHengSeiberg, Lake1d, Altman1d, LRLifshitz}.

This more familiar low-energy universal description naturally emerges
from a more ``microscopic'' (low-energy equivalent) formulation in
terms of a phonon (layer displacement) ${\bf u} = u(\rv)\hat{\bf y}$
and the unit-normal (layer orientation)
$\hat{\bf n}(\rv)=-\hat{\bf x}\sin\hat\theta + \hat{\bf
  y}\cos\hat\theta \equiv \hat{\bf y} + \delta\hat{\bf n}$ field
operators, and the corresponding canonically conjugate linear and
angular momentum fields, $\pi(\rv)$ and $L(\rv)$, with the Hamiltonian
density,
\begin{equation}
  \mathcal{H}_{\text{sm}}=\frac{1}{2}{\bf \pi}^2
  +\frac{1}{2} L^2+\frac{1}{2}\kappa(\grad u+\delta\hat{\bf
    n})^2+\frac{1}{2} K(\grad \hat{\bf n})^2,
\end{equation}
where $\kappa,K$ are elastic constants.

Working in a phase-space path-integral formulation, the corresponding
Lagrangian density is,
\begin{eqnarray}
  \mathcal{L}_{\text{sm}}&=&\pi \p_0 u+L\p_0
  \theta-\frac{1}{2}\pi^2-\frac{1}{2}L^2+\frac{1}{2}\kappa^{-1} {\bf
    \sigma}^2+\frac{1}{2} K^{-1} {\tau}^2\nonumber\\
  &-&{\bm\sigma} \cdot \left(\grad u - \hat{\bf x}\theta\right)
  -{\tv} \cdot \grad\theta\;.
\label{Lsm}
\end{eqnarray}
In above, we neglected $\theta$ nonlinearities, took the x-axis to be
along the smectic layers, and, as with a crystal duality of previous
subsection, for later convenience introduced Hubbard-Stratonovich
fields ${\bm\sigma}$ and $\tv$, corresponding to the local stress and
torque, respectively.  Integrating over the auxiliary fields
$\pi, L, {\bm\sigma},\tv$ easily recovers the phonon-only Lagrangian
in Eq.\rf{LsmLifshitz}.

The form \rf{Lsm} allows us to separate Goldstone modes into smooth
and singular (defects) components and to functionally integrate over
the smooth, single-valued parts of the phonon $u$ and orientation
$\theta$ fields. With this, we obtain coupled linear and angular
momenta conservation constraints,

\begin{eqnarray}
\p_0 \pi-\grad \cdot {\bm\sigma}=0,\;\;
\p_0 L-\grad \cdot \tv=\hat{\bf x}\cdot{\bm\sigma},
\label{momentumContinuity}
\end{eqnarray}
Solving these in terms of gauge fields,
\begin{eqnarray}
\pi&=&\hat{\bf z} \cdot \left( \grad \times {\bf A}\right),\;\;
{\bm\sigma}=\hat{\bf z}\times(\p_0{\bf A}+\grad A_0),\\
L&=&\hat{\bf z} \cdot \left(\grad \times {\bf a}-\hat{\bf x} \times {\bf A} \right),\;\;
\tv =\hat{\bf z}\times(\p_0{\bf a}+\grad a_0 - \hat{\bf x}A_0).
\nonumber
\end{eqnarray}
allows us to express smectic's Lagrangian density in terms of these
Goldstone-mode encoding gauge fields, and to obtain the Maxwell part
of the smectic dual Lagrangian,
\begin{eqnarray}
\tilde\cL^{\text{sm}}_{\text{M}}&=&\frac{1}{2\kappa} \left( \p_0 {\bf
      A}+\grad A_0\right)^2-\frac{1}{2}\left(\grad \times {\bf A}
  \right)^2\\
  &+&\frac{1}{2K} \left(\p_0 {\bf a}+\grad
    a_0-A_0 \hat{\bf x}\right)^2-\frac{1}{2}\left(\grad \times {\bf
      a}-\hat{\bf x}\times {\bf A}\right)^2,\nonumber
\end{eqnarray}
Similarly to crystal's gauge-dual of previous subsection, the
smectic's gauge dual displays a nontrivial ``minimal'' coupling
between the translational and orientational gauge fields, and exhibits
a generalized gauge invariance under transformations,
\begin{subequations}
\begin{eqnarray}
  &&(A_0, {\bf A}) \to A_{\mu}'=\left(A_0-\p_0\phi, 
    {\bf A}+\grad\phi\right),\\
  &&(a_0,{\bf a}) \to a_{\mu}'=\left(a_0-\p_0\chi,
    {\bf a}+\grad\chi-\hat{\bf x}\phi\right).\;\;\;\;\;\;
\end{eqnarray}
\label{gauge_transform}
\end{subequations}
The six gauge field degrees of freedom $A_\mu, a_\mu$ reduce to two
physical Goldstone modes after gauge fixing $\phi,\chi$ and
implementing two Gauss's law constraints \rf{GaussLaws}.

To include dislocations and disclinations we allow for the
nonsingle-valued component of $u$ and $\theta$, respectively defined
by
\begin{subequations}
\begin{eqnarray}
  p&=&\hat{\bf z}\cdot\grad\times\grad u,\;\;\;
  {\bf J} =\hat{\bf z}\times\left(\grad\p_0 u
-\p_0\grad u\right),\\
  \rho&=&\hat{\bf z}\cdot\grad\times\grad\theta,\;\;\;
  {\bf j}=\hat{\bf z}\times\left(\grad\p_0 \theta
    -\p_0\grad\theta \right).
\end{eqnarray}
\end{subequations}
This together with $\mathcal{\tilde L}^{\text{sm}}_{\text{M}}(A_\mu,a_\mu)$
gives the dual Lagrangian density for the quantum smectic,
\begin{equation}
  \tilde{\mathcal{L}}_{\text{sm}}=\mathcal{\tilde L}^{\text{sm}}_{\text{M}}(A_\mu,a_\mu) 
  + {\bf A} \cdot {\bf J} - A_0 p
  + {\bf a} \cdot {\bf j} - a_0 \rho,
\label{dualLsm}
\end{equation}
corresponding to the Hamiltonian density
\begin{equation}
\begin{split}
  \tilde{\mathcal{H}}_{\text{sm}}=&\frac{1}{2}\kappa{\bf
    E}^2+\frac{1}{2}(\grad \times {\bf A})^2+\frac{1}{2}K
  {\bf e}^2\\
  &+\frac{1}{2}( \grad \times {\av}-\hat{\bf x} \times {\bf
    A})^2-{\bf A} \cdot {\bf J}-{\bf a} \cdot {\bf j},
\end{split}
\label{dualHsm}
\end{equation}
supplemented by the generalized Gauss's law constraints,
\begin{eqnarray}
  \grad \cdot {\bf E}&=&p-{\bf e}\cdot \hat{\bf x},\;\;\;
  \grad \cdot {\bf e}=\rho.
\label{GaussLaws}
\end{eqnarray}
$p$ and ${\bf J}$ are $\hat{\bf x}$-dipole charge and current
densities, representing $\hat{\bf y}$-dislocations, while $\rho$ and
${\bf j}$ are fractonic charge and current densities, corresponding to
disclinations. The generalized gauge invariance of \rf{dualHsm}
imposes coupled continuity equations for the densities
\begin{equation}
\p_0 p + \grad \cdot {\bf J}=-{\bf j}\cdot\hat{\bf
    x}\;,\ \ \ \ \ \ \ 
  \p_0\rho + \grad \cdot {\bf j} = 0,
\label{smecticContinuity}
\end{equation}
where dipole conservation is violated by a nonzero charge current,
${\bf j}\cdot\hat{\bf x}$ along smectic layers.

\begin{center}
\begin{figure}[htbp]
\centering
 \hspace{0in}\includegraphics*[width=0.48\textwidth]{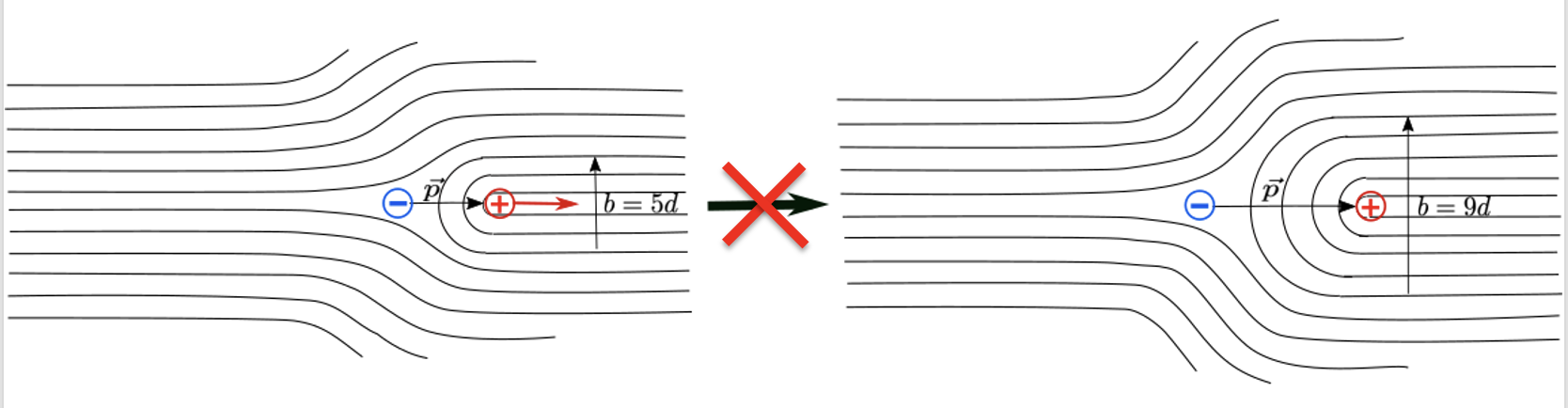}
 \caption{Illustration of restricted along-layers mobility of $+/-$
   disclination (lineon) charges (making up a dislocaton $b$, i.e., a
   dipole $p$) in a quantum smectic, forbidding their separation, that
   corresponds to a nonlocal process of adding a smectic half-layer
   per lattice constant of charge separation.}
 \label{mobilitySmDisclinationFullFig}
\end{figure}
\end{center}

These equations thus transparently encode the restricted mobility of
the disclination charges $\rho$, illustrated in
Fig. \ref{mobilitySmDisclinationFullFig}, with a relaxation rate
$\Gamma_k = D k_y^2 + \gamma k_x^4 $, resulting in slow subdiffusive
decay $\rho(t)\sim 1/t^{3/4} $, and mobility and diffusion coefficient
vanishing along the $\hat{\bf x}$-directed smectic layers, i.e.,
${\bf j}\cdot\hat{\bf x} = 0$ in the absence of dislocation dipoles
\cite{LRsmecticPRL2020, Feldmeier_anomalous, subdiffusionPrinceton, GLN_hydro, glorioso_hydro, GGL_hydro}.


\subsection{Quantum melting}
\label{ss:melt}

Starting with the gauge-dual of the quantum crystal (derived in the
earlier subsections \rf{sec:ElasticDuality}), either in tensor, \rf{tensorGLcrystalDual}
or in its equivalent coupled-vector gauge theory form, \rf{LdualCr},
and minimally coupling it to dynamic dislocation-dipole matter,
$\psi_d$, gives a quantum Lagrangian density,
\begin{eqnarray}
\hspace{-0.2cm}\tilde{\mathcal{L}}_{\text{cr}}&=&\sum_{k = 1,2}
\frac{J_k}{2}|(\partial_\mu - i A^k_\mu)\psi_k|^2 - V(\{\psi_k\})\nonumber\\
&&+ \mathcal{L}^{\text{cr}}_{\text{M}}(A_{1,\mu},A_{2,\mu},a_\mu),
\label{dualLcrDislocations}
\end{eqnarray}
where $\psi_{k=1,2}$ correspond to $\hat{\bf p}_{1,2}$ dipoles (two
elementary dislocations), $A_{k,\mu},a_\mu$ gauge fields capture the
$k$-th phonons and bond orientational Goldstone modes (with unit
dipole-charges $p_k$ absorbed in the definitions of $A_{k,\mu}$), and
$V(\psi_1,\psi_2)$ a $U(1)$ symmetric Landau potential, with a dual
Maxwell Lagrangian \rf{LdualCr}.  This field theory\footnote{As
  discussed in the following subsection, for a complete description it
  must be supplemented with vacancies/interstitials (atoms) sector,
  captured by a conventional Abelian-Higgs model with bosonic matter.}
thus gives a complete description of the quantum melting transitions
out of the fractonic crystal state.  The nature of such quantum
transition is then dictated by the form of the dipole interaction
potential, $V(\psi_1,\psi_2)$.

\subsubsection{Crystal-to-hexatic Higgs transition}

In a 2d hexagonal (square) crystal, the $C_6$ ($C_4$) invariance
enforces the symmetry between $\psi_1$ and $\psi_2$ dislocation
dipoles, i.e., their interaction potential $V(\psi_1,\psi_2)$ is
$1\leftrightarrow 2$ symmetric.  Thus, as illustrated in
Fig. \ref{crystalHexaticFig}, driven by quantum fluctuations, both
types of dipoles unbind and Bose-condense at a single Higgs
transition, that is the quantum crystal-to-hexatic (tetratic) melting,
the counterpart of the famous 2d classical thermal melting transition,
predicted by Kosterlitz-Thouless, Halperin-Nelson, and
Young\cite{KT,HN,Young, Z3LRsineGordonPRB}.\footnote{A classical
  electrostatic limit of the crystal's gauge dual reduces
  \rf{dualLcrDislocations} to a vector sine-Gordon model
  $\tilde H = \frac{1}{2} \tilde{C}^{-1}_{ij,kl} \partial_i \partial_j
  \phi \partial_k\partial_l \phi -g_b \sum_{n=1}^{p} \cos({\bf b}_n
  \cdot \hat{z} \times \nabla \phi) -g_s \cos(s_p\phi)$ that
  reproduces\cite{Z3LRsineGordonPRB} the two-stage
  crystal-hexatic-isotropic melting of KTHNY.} The Higgs transition
thus gaps out {\em both} translational gauge fields, $A_{1,\mu}$ and
$A_{2,\mu}$, which can therefore be safely integrated out, or to
lowest order effectively set to zero. This reduces the coupled gauge
theory to a conventional Maxwell form for the remaining rotational
gauge field $a_\mu$, with
\begin{equation}
  \mathcal{L}^{\text{hex}}_{\text{M}}(a_\mu)\approx
  \mathcal{L}^{\text{cr}}_{\text{M}}(A_{1,2\mu}\approx 0, a_\mu)=
  \frac{1}{2}K^{-1}{\bf e}^2 - \frac{1}{2}(\grad \times {\bf a})^2.
  \label{CrToHexatic}
\end{equation}
As expected it is the dual to the quantum XY-model of the
orientationally-ordered quantum hexatic (tetratic)
liquid\footnote{Because, a condensation of dislocations necessarily
  leads to Bose-condensation of bosonic vacancies and interstitials,
  the resulting hexatic fluid is necessarily a superfluid, i.e., a
  superhexatic.},
$\mathcal{L}_{\text{hex}} =\frac{1}{2}(\p_0\theta)^2 -
\frac{1}{2} K(\grad\theta)^2$. As with the conventional U(1) Higgs
(normal-superconductor) transition, mean-field approximation breaks
down for $d+1\leq 4$, and may be driven first-order by translational
gauge-fields $A_{k,\mu}$ fluctuations.\cite{HLM,LRscsa} Analysis of
the non-mean-field criticality of this quantum crystal-superhexatic
(supertetratic) melting transition remains an open problem.

\subsubsection{Crystal-to-smectic Higgs transition}

\begin{figure}[htbp]
  \hspace{-0.13in}\includegraphics*[width=0.5\textwidth]{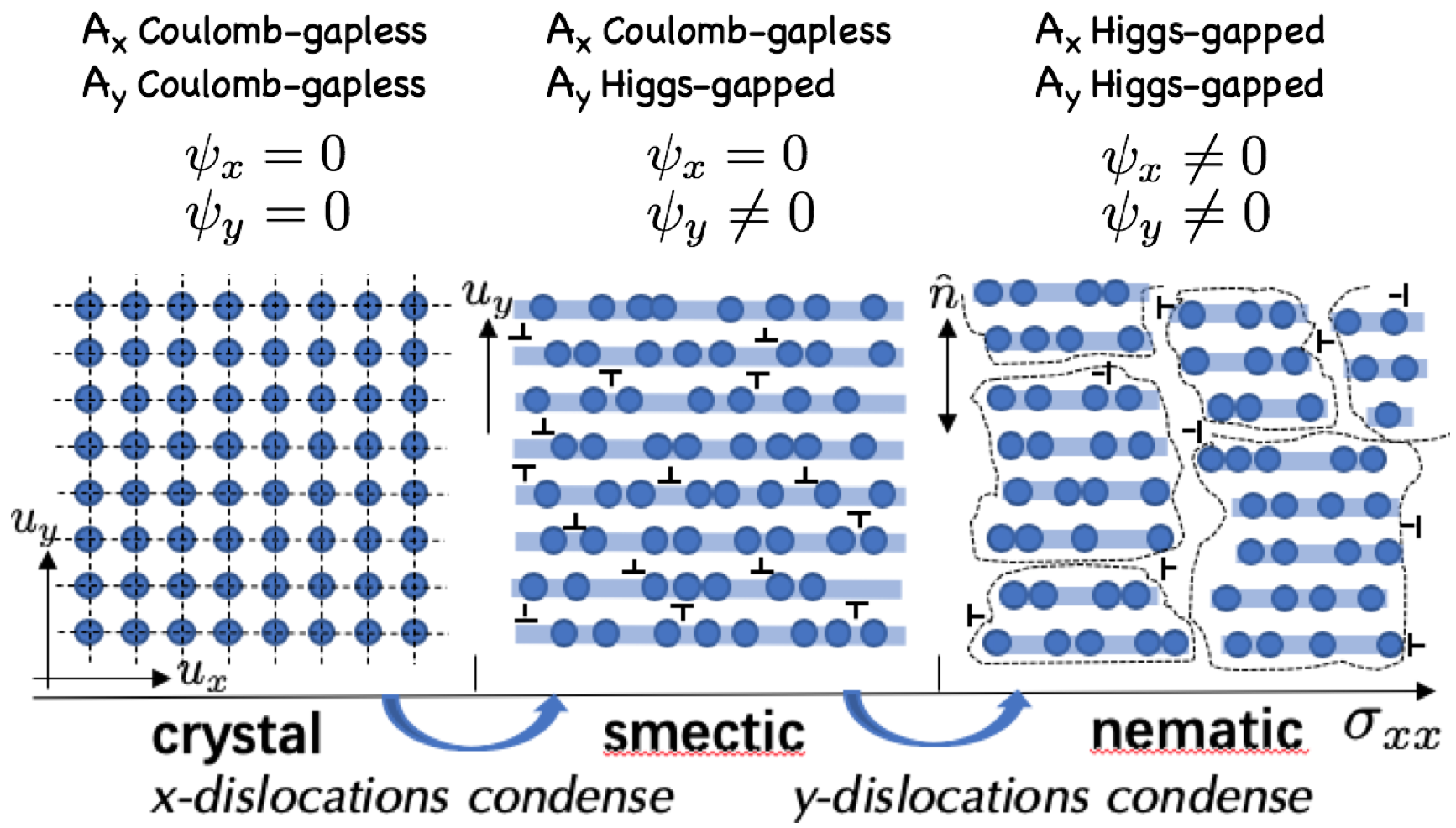}
  \caption{Illustration of quantum melting of a 2D crystal into a
    smectic, followed by smectic-to-nematic melting, respectively
    driven by a condensation of $b_x$ ($\psi_y$ dipoles) and of $b_y$
    ($\psi_x$ dipoles) dislocations.}
  \label{phaseDiagramCrystSmectNemOrdersFig}
\end{figure}

An alternative to the above {\em direct}, continuous
crystal-to-hexatic (or to-tetratic) quantum melting
scenario is a {\em two-stage} transition that takes place in a
uniaxial crystal where $C_6$ or $C_4$ symmetry is broken down to
$C_2$, as illustrated in Fig. \ref{phaseDiagramCrystSmectNemOrdersFig}.
The reduced uniaxial symmetry is necessarily encoded in the Landau
potential $V(\psi_1,\psi_2)$, specifically, controlled by the
quadratic term $g_k|\psi_k|^2$, with $g_2 < g_1$ leading to a
condensation of $\psi_2$ dislocation dipoles, with $\psi_1$ remaining
gapped.

Alternatively, this breaking of $C_6$ (or $C_4$) symmetry down to
$C_2$ may happen spontaneously, with $g_1 = g_2$, instead controlled
by the sign of the $v$ coupling in the dislocation interaction,
$v |\psi_x|^2|\psi_y|^2$.  For $v > 0$, only one of the two dipole
species condenses, say $\psi_2\neq 0$ with $\psi_1 = 0$. This Higgs
transition thus only gaps out $A_{2,\mu}$, which then can be safely
integrated out. To lowest order this corresponds to
$A_{2.\mu}\approx 0$, reducing crystal's Maxwell Lagrangian to that of
the quantum smectic \rf{LsmLifshitz}, with
\begin{eqnarray}
  \mathcal{L}^{\text{sm}}_{\text{M}}(A_{1,\mu}, a_\mu)\approx
  \mathcal{L}^{\text{cr}}_{\text{M}}(A_{1,\mu},A_{2,\mu} \approx 0, a_\mu).
\end{eqnarray}

\begin{center}
\begin{figure}[htbp]
\centering
 \hspace{0in}\includegraphics*[width=0.48\textwidth]{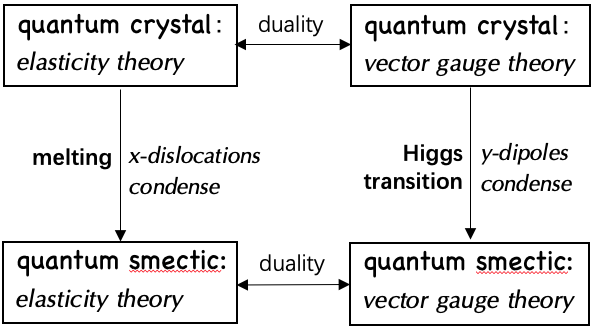}
 \caption{Quantum crystal-smectic duality relations and the associated
   quantum melting transition.}
\label{crystalTOsmecticFig}
\end{figure}
\end{center}

While this phase transition faithfully captures the crystal-smectic
melting at the mean-field level, as with the crystal-hexatic Higgs
transition, its true critical properties are expected to be nontrivial
and remain to be analyzed.  The two ways of obtaining the smectic
gauge dual -- by dualizing a smectic Lagrangian and through the above
Higgs melting transition of the crystal's gauge dual -- are summarized
in Fig.\ref{crystalTOsmecticFig}.

\subsubsection{Quantum smectic-to-nematic Higgs transition}

As illustrated in Fig. \ref{phaseDiagramCrystSmectNemOrdersFig}, the
above crystal-to-smectic transition is then naturally followed by
quantum melting into a nematic superfluid by condensation of
$\psi_1$-dipoles (aligned with the smectic layers), i.e., a
proliferation of $\bv_1$ dislocations with Burgers vectors
transverse to smectic layers.  The resulting $\psi_1\neq 0$ Higgs
phase gaps out the remaining smectic translational gauge field
$A_\mu (=A_\mu^x$), which can therefore be safely integrated out. This
reduces the model to a conventional Maxwell form for the rotational
gauge field $a_\mu$, with
\begin{equation}
\mathcal{L}^{\text{nem}}_{\text{M}}(a_\mu)\approx
\mathcal{L}^{\text{sm}}_{\text{M}}(A_\mu \approx 0, a_\mu)=
\frac{1}{2}K^{-1}{\bf e}^2 - \frac{1}{2}(\grad \times {\bf a})^2,
\label{smecticTOnematic}
\end{equation}
that is a dual to the quantum XY-model of the nematic,
$\mathcal{L}_{\text{nem}} =\frac{1}{2}(\p_0\theta)^2 -
\frac{1}{2} K(\grad\theta)^2$. Fluctuation corrections lead to an
anisotropic stiffness and subdominant higher order gradients. As with
the conventional U(1) Higgs (normal-superconductor) transition,
mean-field approximation breaks down for $d+1\leq 4$, and may be
driven first-order by translational gauge-field, $A_\mu$
fluctuations.\cite{HLM, LRscsa} Analysis of the non-mean-field criticality of
the quantum smectic-nematic transition also remains an open problem.

\subsection{Supersolid, superhexatic, supersmectic: vacancies and interstitials}
\la{supersolid}

\begin{figure}[htbp]
  \hspace{0in}\includegraphics*[width=0.25\textwidth]{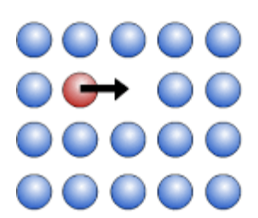}
  \caption{Mobile vacancy (and interstitial) defects of a crystal,
    necessary to faithfully capture the associated supersolid and
    superfluid phases.}
\label{vacancyMobileFig}
\end{figure}

\noindent{\em Vacancies and interstitials.} As discussed in
Sec.\ref{sec:ElasticDuality}, so far we have neglected vacancies and
interstitials, (see Fig. \ref{vacancyMobileFig}), which physically
corresponds to a restriction to their Mott-insulating,
commensurate-crystal state.  This clearly misses the additional atomic
sector of the system, encoded by a Bose-Hubbard (quantum XY) model or
its gauge-dual Abelian Higgs model from Sec.\ref{PVD}. The need for
this missing vacancies/interstitials (atomic) sector is clear as
quantum melting a crystal of bosons, at zero temperature and in the
absence of substrate or disorder generically leads to a gapless
superfluid.
\begin{figure}[t!]
 \centering
 \includegraphics[scale=0.7]{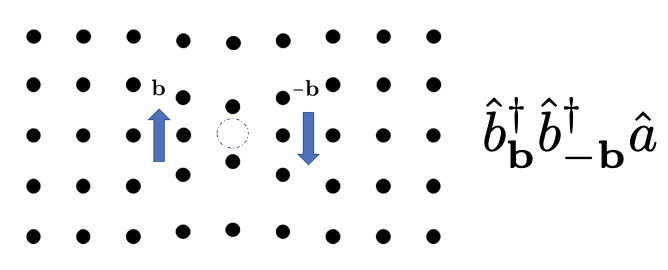}
 \caption{A disclination quadrupole, constructed as a bound state of
   two equal and opposite dislocations with Burgers vectors $\bf b$
   and $-\bf b$, carries a unit of vacancy (atom) number - a local
   defect that can be seen as a deficiency of an atom in the middle of
   the configuration, as illustrated in
   Fig. \ref{vacancyMobileFig}. This construction demonstrates the
   allowed ${\hat b^\dagger}_{\bf b} {\hat b}^\dagger_{\bf b}{\hat a}$
   operator, which encodes that condensation of dipoles is necessarily
   accompanied by condensation of vacancies and intetstitials and thus
   leading to superfluidity of the fluid phases.}
 \label{dipoleVacancySFfig}
\end{figure}

\begin{figure}[t!]
 \centering
 \includegraphics[scale=0.33]{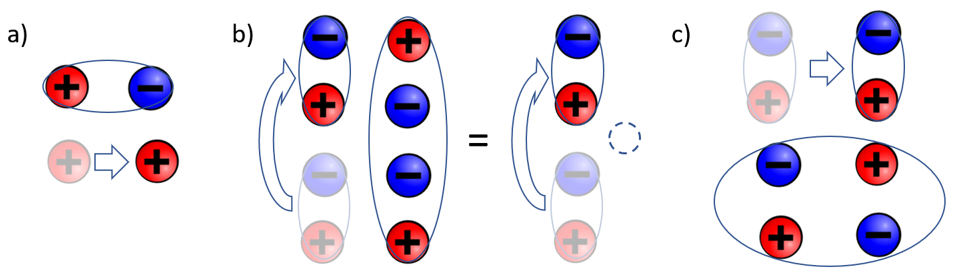}
 \caption{(a) Fracton motion is forbidden as it requires emission of a
   conserved dipole.  (b) In the F$_{U(1)}$ charge enriched fractonic
   phase (commensurate crystal) \emph{longitudinal} dipole motion
   (dislocation climb) is forbidden as it requires emission of a
   linear quadrupole carrying conserved vacancy/interstitial number,
   corresponding to local compression of the crystal, $i.e.$, a
   vacancy defect.  c) \emph{Transverse} dipole motion (dislocation
   glide) is allowed as it creates a square quadrupole, corresponding
   to a local shear.}
 \label{dipoleMobilityFullFig}
\end{figure}

\begin{figure}[t!]
 \centering
 \includegraphics[scale=0.4]{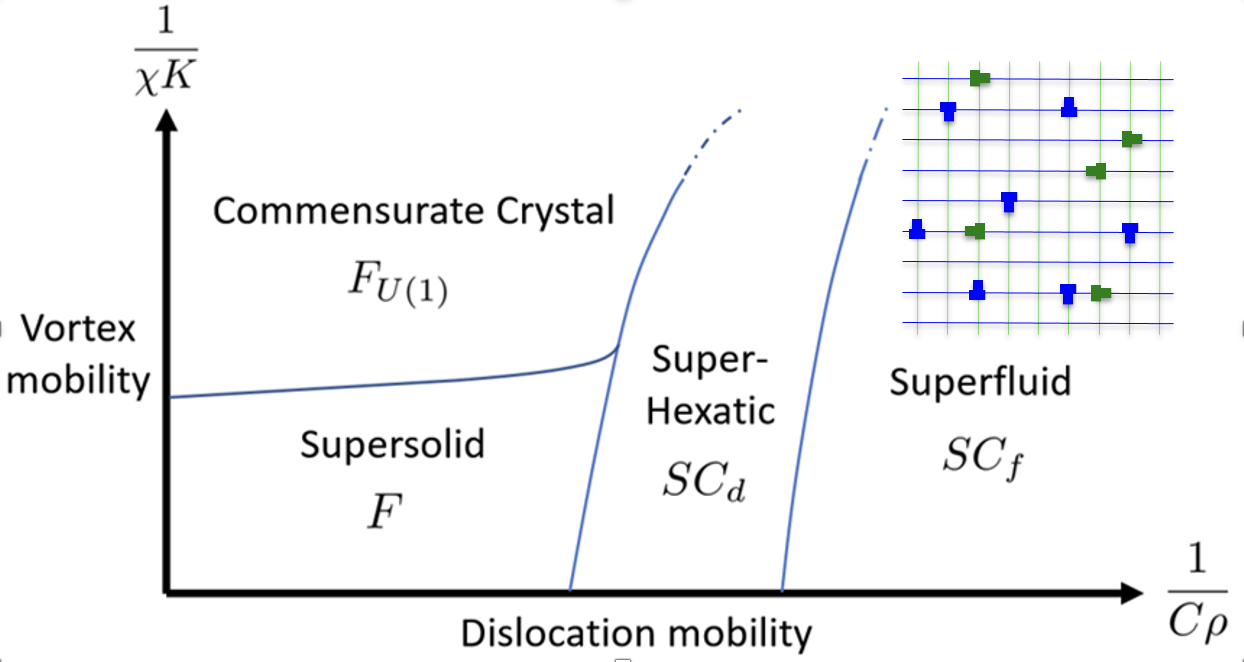}
 \caption{A schematic phase diagram illustrating phases derived from
   the supersolid (a $U(1)$-symmetry broken fracton phase, $F$).  Upon
   condensation of vortex defects, bosons can transition to a
   commensurate crystal (a $U(1)$-symmetric fracton phase,
   $F_{U(1)}$), super-hexatic, or superfluid phase.  Note that
   $U(1)$-symmetric liquid and hexatic phases are forbidden at zero
   temperature for reasons summarized in
   Fig. \ref{dipoleVacancySFfig}.}
 \label{crystalHexaticFig}
\end{figure}

Thus, as discussed in detail in
Refs.\onlinecite{PretkoLRsymmetryEnrichedPRL2018,PretkoZhaiLRdualityPRB2019,KumarPotter19}
and illustrated in the phase diagram of Fig. \ref{crystalHexaticFig},
at zero temperature two qualitatively distinct - commensurate and
incommensurate quantum crystals are possible, respectively
distinguished as Mott-insulating and superfluid states of this atomic
(vacancies/interstitials) sector. In the former, the U(1)
symmetry-enriching constraint imposes a glide-only motion of
dislocation (illustrated in Figs. \ref{climbGlideFig} and
\ref{dipoleMobilityFullFig}), that is broken in the latter, where
dipole dislocation motion is
unconstrained.\cite{MarchettiLR1999, PretkoLRsymmetryEnrichedPRL2018}

Under duality, these two types of quantum crystals then map onto two
distinct fractonic phases, F$_{U(1)}$ and F, respectively, with and
without quadrupole-imposed restriction on the dipole glide-only and
unrestricted motion, as illustrated in
Fig. \ref{dipoleMobilityFullFig}(b,c). \cite{PretkoLRsymmetryEnrichedPRL2018}

This dipole constrained motion condition is concisely encoded in the
Ampere's law of the corresponding tensor gauge theory,
\begin{equation}
  \p_0 E^{ij} + \frac{1}{2}(\epsilon^{ik}\partial_kB^j +
  \epsilon^{jk}\partial_kB^i) = -J^{ij}\;,
\end{equation}
whose trace
can be expressed in terms of vacancies-interstitials density
$n_d = E^i_{\,\,i} + n_0\p_i u_i$ and the corresponding current,
$ J^i_{d} = \pi^i$,
\begin{equation}
\p_0 n_d + \partial_i\pi^i = -J^i_{\,\,i}
\label{continuity}
\end{equation}
where we used $E^i_{\,\,i} = n_d - n_0\p_i u_i\approx n_d$, first
derived in Ref. \onlinecite{MarchettiLR1999}.
With this vacancies-interstitials continuity equation (sourced by
$J^i_{\,\,i}$, corresponding to the longitudinal (along-dipole) motion
of dipoles \cite{genemPretko, PretkoLRsymmetryEnrichedPRL2018}),
Eq. \ref{continuity} encodes that climb of dislocations creates
vacancy/interstitial defects.  Conversely, Mott-insulating
(commensurate-crystal) state of the latter, restricts the motion of
dislocations to glide-only lineon type, dualizing to a
symmetry-enriched fracton state F$_{U(1)}$. It can then undergo a
quantum phase transition to a distinct fracton state F, when vacancies
and interstitials Bose-condense, thereby breaking the atom-number
$U(1)$ symmetry. The corresponding phase diagram is illustrated in
Fig.\ref{crystalHexaticFig}.

In contrast to a crystal, that allows F$_{U(1)}$ ground state, we
observe that hexatic and smectic states are dislocation condensates
(corresponding to condensation of both or just one of the $\hat\xv-$
and $\hat\yv-$dislocation dipoles [created by
$\hat{b}^\dagger_{\bf b}$]). Thus, vacancies and interstitials
(created by $\hat{a}^\dagger$, illustrated in
Fig.\ref{dipoleVacancySFfig}), consisting of pairs of
oppositely-charged dislocations (disclination charge quadrupoles), are
necessarily driven to Bose-condense by the allowed coupling
$\hat{a} \hat{b}^\dagger_{\bf b}\hat{b}^\dagger_{-\bf b}$. Thus, a
hexatic and a smectic are necessarily incommensurate ``super-hexatic''
and ``super-smectic'', respectively, and their dual gauge theory
\rf{dualHsm} is implicitly understood to be coupled (via axion-like
${\cal E}-B$, ${\cal B}-E$ couplings) to a conventional $U(1)$ gauge
theory with fields ${\cal E, B}$ -- a dual to the liquid of vacancies
and interstitials.\cite{PretkoLRsymmetryEnrichedPRL2018,
  PretkoZhaiLRdualityPRB2019}

\subsection{External perturbations}

With an eye to experimental probes, we briefly discuss the role of
external perturbations.  Crystal's analog of a chemical potential is
the imposed velocity -- a ``momentum chemical potential'', which
imposes a nonzero density of finite-momentum boson density,
$n_{\bf G}$, i.e., a nonzero momentum on the crystal. On the dual side
this corresponds to an external dual-magnetic field. In the
dual-nonsuperconducting state of vanishing disclination and
dislocation density, the response is linear, as expected by crystal's
Galilean invariance. In contrast, the dual-superconductor expels the
imposed flux, either completely in the dual-Meissner or ``mixed''
Abrikosov-like state, respectively corresponding to a viscous response
of a fluid or as a lattice of dislocations, carrying a nonzero
momentum.

A crystal can also be subjected to a compressive or shear stress,
which on the dual side is an imposition of a tensor electric field.  A
dislocation will generically be set in motion by the associated
Peach-Koehler force, $E_{ij} p_j$ encoded as an imposed dual
electrictrostatic field on a charged dipole particle, $p_j$.  In the
absence of dipoles this probes the response of external tensor field
across the dual ``dielectric''. For stress above a critical value, a
dual dielectric breakdown will take place, corresponding to a
proliferation of dipole dislocations.  The response of a smectic is
more complex -- it is dual Meissner-like along and non-superconducting
across the smectic layers.

A substrate also plays a qualitatively interesting role, as it breaks
the underlying rotational and translational symmetries, thereby
breaking angular and linear momentum conservations.  Repeating the
duality analysis for a translationally incommensurate substrate, we
find that, it reduces the orientational $U(1)$ sector $(\ev, \av)$ to
a discrete $Z_n$ gauge theory (for $n$-fold orientational
commensurability), coupled to a noncompact $U(1)$ translational-sector
gauge theory $(\E_k, \A_k)$. For $n=1$ a la Polyakov confinement in
2+1d, the orientational degree are eliminated and the translational
sector reduces to two conventional decoupled $U(1)$ gauge theories,
for $k=x, y$. These are compact in the presence of a translationally
$p$-fold commensurate substrate and will also be reduced to discrete
$Z_p$ gauge theories, and fully confined, i.e., pinned for $p=1$.

\subsection{Vortex crystal }
We now consider constrained dynamics of vortices. The first system of
interest is a vortex crystal. We assume that it was formed by
nucleating a large number of vortices in a superfluid. The elasticity
of a vortex crystal is described in terms of both superfluid phase
degree of freedom and vortex-lattice phonons. The low energy
Lagrangian takes the form
\be\la{eq_lag_vc}
\mathcal L = - \frac{1}{2}\Gamma\bar{n} \epsilon^{ij} u_i \p_0u_j - \frac{1}{2}C_{ijkl} u_{ij}u_{kl} + \Gamma e_i u^i + \frac{1}{g^2} \left(e^2 - b^2\right)\,,
\ee
where $\Gamma$ is the vorticity, $\bar{n}$ is superfluid density, and
dual electromagnetic fields, $e_i$ and $b$ capture the Goldstone modes
of the superfluid (see Section \ref{PVD}). The first term (single time
derivative Berry's phase-like term) in \eqref{eq_lag_vc} is unique to
the vortex crystals and explicitly breaks parity. It originates from
the Magnus force (associated with a nonzero boson density -- seen by
vortices as an effective magnetic field) experienced by vortices,
encoding noncommutativity of $u_x$ and $u_y$.

The duality transformation follows the steps similar to the previous
sections and the final dual Lagrangian is \cite{NGM_chiral,
  PretkoZhaiLRdualityPRB2019}
\begin{multline}  \la{eq_dlag_vc}
\tilde{\mathcal L} = \frac{1}{2\Gamma \bar{n}} \epsilon^{ij}\left(B_i - \Gamma \epsilon_{i}{}^k a_k \right)\p_0\left(B_j - \Gamma \epsilon_{j}{}^l a_l \right) 
\\
 + \frac{1}{2} \tilde{C}^{-1}_{ijkl} \left(E^{ij} + \Gamma \delta^{ij} a_0 \right) \left(E^{kl} + \Gamma \delta^{kl} a_0\right)\,.
\end{multline}
The tensor and vector gauge sectors are coupled through a non-trivial
minimal-like coupling akin to the coupled vector gauge theories of a
crystal, Sec. \ref{coupledVector} and smectic, Sec. \ref{smecticDual}
duals.\cite{RHvectorPRL2020, LRsmecticPRL2020, ZRsmecticAOP2021} The
gauge transformations act as follows
\be \delta A_{ij} = \p_i \p_j
\alpha + \Gamma\delta_{ij}\,,\quad \delta A_0 = - \p_0\alpha\,,\quad
\delta a_\mu = \p_\mu \beta\,.
\ee
The first term in
\eqref{eq_dlag_vc} has no analogue in ordinary electro-magnetism. The
dipole conservation law \eqref{eq_elast} remains the same, however the
glide constraint is modified due to the possibility of vortex creation
in the superfluid, akin to earlier discussion of vacancies and
interstitials in an incommensurate atomic crystal,
\rf{continuity},
\be\la{eq_glide2}
\p_\mu j^\mu =  J^{ii}\;,
\ee
where $j^\mu$ is the superfluid vortex current. The glide constraint
\eqref{eq_glide2} states that the dislocations can climb at the
expense of creating vortices (\emph{i.e.}, violating vortex continuity
equation $\p_\mu j^\mu = 0$).\cite{MarchettiLR1999}

Detailed analysis of vortex lattice melting, similar to the discussions of Sect. \ref{ss:melt}, was presented in \cite{NGM_chiral}.

\subsection{Vortex fluid}
Remarkably, a classical system of interacting vortices (or,
equivalently electric charges in a strong magnetic field)
conserves dipole moment on its own, and is thus fractonic. We
demonstrate this on the example of a classical system of $N$ vortices.

On increasing vortex density (e.g., by rotation), we expect a vortex
crystal to melt into a vortex fluid.  Neglecting dissipation, at zero
temperature a vortex system can be approximated as Hamiltonian for
any number of vortices and is described by the following Lagrangian
\be\la{eq_vfluid}
\mathcal L = 2\pi\sum_\alpha \gamma_\alpha x_1^\alpha \p_0x_2^\alpha  - 2\pi \sum_{\alpha < \beta} \gamma_\alpha \gamma_\beta \ln |x^\alpha - x^\beta|\,.
\ee
We note that this Lagrangian neglects effects of vortex drag and the normal component of a superfluid.

Due to translational and rotation invariance, the total linear and
angular momenta are conserved. However, due to non-commutative nature
of vortex coordinates in \eqref{eq_vfluid} the linear momentum is
equal to the dipole moment rotated by $\frac{\pi}{2}$ and angular
momentum coincides with the trace of the quadrupole moment
\be
P_i = \epsilon_{ij} \sum_\alpha
\gamma_\alpha x^\alpha_j = \epsilon_{ij} Q_j\,, \quad L = \sum_\alpha
\gamma_\alpha x^\alpha_j x^\alpha_j = \text{tr}(Q_{ij})\,.
\ee
Consequently, a vortex dipole moves perpendicular to the dipole
moment, while isolated vortices are immobile
\cite{DG_vortices}. Exactly the same Lagrangian describes electrons in
the lowest Landau level, where these conservation laws are related to
the area-preserving diffeomorphism symmetry
\cite{GromovDualityPRL2019, DMNS_APD}.

The vortex lattice discussed in the previous section can melt into a
vortex liquid. This liquid can be understood as a hydrodynamic limit
of \eqref{eq_vfluid}. It retains the same conservation laws as the
finite $N$ system and its continuity equation takes the same form as
the traceless scalar charge theory \eqref{eq_elast}.

\subsection{Geometric theory of defects}
In the previous sections we discussed two complementary approaches to
elasticity and crystalline defects. In this section we will make a
connection between fractons and geometric description of crystalline
defects. This description dates back to the work of Kondo \cite{kondo_geo}, and
is valid in all spatial dimensions. Disclinations and dislocations are
described using Riemann-Cartan (RC) geometry, while the phonons are
described by the fluctuations of metric. The geometric theory of
defects leverages RC geometry to describe physical properties of
defects as well as defect-phonon scattering \cite{KV_geo}.

The description of dislocations and disclinations in RC geometry can
be understood by noting that torsion, $T^a_{ij}$ and curvature
$R^{ab}_{ij}$ -- the main ingredients of the RC geometry -- correspond
to defects in translational and rotational symmetries. This follows
directly from the definition\footnote{Here indices $a,b,c,\ldots$
  refer to the tangent space, while indices $\mu,\nu,\rho,\ldots$
  refer to the spacetime and indices $i,j,k,\ldots$ refer to the space.}
\be\la{eq_RT}
[\nabla_i, \nabla_j] v^a = T^b_{i j} \partial_b v^a + R^{a}{}_b{}_{;ij} v^b\,.
\ee
Equation \eqref{eq_RT} states that transporting a vector $v^a$ around
a small loop leads to an infinitesimal rotation by $R^{ab}$ (Franck
angle) and translation by $T^a$ (Burgers vector).

The relation between dislocations and disclination dipoles is built into the structure of RC geometry and is phrased as a relation between the Levi-Civita curvature and torsion
\be\la{eq_RLCT}
2 R = \p_i \Big(\epsilon^{a}{}_b  e^i_a T^b\Big)\,,
\ee
where $e^i_a$ is the frame field, $T^b = T^b_{ij} \epsilon^{ij}$ and $R$ is the Ricci scalar curvature constructed from the curvature two-form $R^{a}{}_b{}_{;ij}$ \cite{Nakahara}. A similar relation plays the foundational role in the teleparallel formulation of gravity, in which the spacetime geometry is described using torsion \cite{BH_gravity}. RC geometry further supplies us with a geometric formulation of \eqref{eq_dipolecontinuity}, \eqref{ContinuityCr}, \eqref{smecticContinuity} by the virtue of the Bianchi identity. We illustrate the relationship in two spatial dimensions. It becomes physically transparent when we define the dislocation and disclination currents in terms of torsion and curvature according to
\be\la{eq_dcurrent}
J^\mu_{a} = \epsilon^{\mu\nu\rho}  T_{a,\nu\rho}\,, \qquad \Theta^\mu = \epsilon^{\mu\nu\rho} R_{\nu\rho}\,.
\ee
Then the Bianchi identity takes form \eqref{eq_dipolecontinuity}
\be\la{eq:discldisl}
\nabla_\mu J^\mu_a = \epsilon_{ab} e^b_\rho \Theta^\rho\,.
\ee
Finally, we should discuss the origins of the glide constraint in the RC language. The glide constraint requires extra information regarding the conservation of total number of lattice sites. The latter can be formulated in an elegant geometric way as follows. First we introduce a current of lattice sites as follow
\be
J^\mu = \epsilon^{\mu\nu\rho}\epsilon_{ab} e^a_\nu e^b_\rho\,.
\ee
Then the conservation of the number of lattice sites takes the form of continuity equation
\be\la{eq_RCglide}
\p_\mu J^\mu = 0\,. 
\ee
The conserved quantity is the total volume, which translates to the total number of lattice sites
\be
V = \int d^2 x J^0 = \int d^2x \det(e^a_i)\,.
\ee
The glide constraint becomes more transparent after writing out \eqref{eq_RCglide} as
\be
\p_0 \det(e^a_i) + 2 e^b_0 \epsilon_{ab} J^a_0 = 0\,,
\ee
where $J^a_0 = \sum_I b^a \delta(x-x_I)$ is the dislocation density defined by \eqref{eq_dcurrent} and temporal frame $e^b_0$ plays the role of velocity field. Thus local volume $\det(e^a_i)$ changes when the dislocations are carried in the direction perpendicular to the Burgers vector. Eq.\eqref{eq_RCglide} has to be postulated in addition to the RC structure.

\subsection{Diverse realizations of tensor gauge theories}
Since original identification of fractons with crystals and liquid
crystals \cite{LRsmecticPRL2020,
ZRsmecticAOP2021,
PretkoLRdualityPRL2018,
PretkoLRsymmetryEnrichedPRL2018,
PretkoZhaiLRdualityPRB2019,
KumarPotter19,
GromovDualityPRL2019,
Z3LRsineGordonPRB,
RHvectorPRL2020,
GS_cosserat,
Gromov_smectic,
NGM_chiral},
fractons have naturally appeared in a number of other elastic systems
that support geometric defects. Again, fractons emerge after an
appropriate duality transformation.  Here we review a few of these
interesting connections.

\subsubsection{Fragile amorphous solids}

A symmetric tensor gauge theory and its associated fractonic order
also recently found application in amorphous fragile solids and
granular media. These are highly nonequilibrium and heterogeneous
solid states, that can sustain external shear\cite{Jam_review2,
  jam_review}. The effective long wavelength elasticity
\cite{amorphous_fracton} emerges from local force and torque balance
constraints of mechanical equilibrium on every grain when force chain
of contacting grains percolate.  In a continuum these can be encoded
through a condition of mechanical equilibrium on the local symmetric
stress tensor $\sigma_{ij}(\rv)$ and external force $f_i(\rv)$,
satisfying,
\begin{eqnarray}
  \partial_i\sigma_{ij}(\rv) = f_j(\rv).
\label{stressBalanceSolid}
\end{eqnarray}

As with crystalline solids this static equilibrium condition
\rf{stressBalanceSolid} can be naturally interpreted as a generalized
Gauss's law,
\begin{eqnarray}
  \partial_iE_{ij}(\rv) = \rho_j(\rv).
\label{GaussVCT}
\end{eqnarray}
for a vector-charge U(1) rank-2 tensor gauge theory \cite{subPretko},
with a symmetric electric field tensor, $E_{ij}$ and the vector charge
density $\rho_i$ describing external force
$f_i$.\cite{amorphous_fracton} This formulation then automatically
encodes the net force and torque balance through vector charge and
dipole moment neutrality.

The amorphous solid elasticity is then postulated to be governed by
the pseudo-electrostatics, with energy density
${\cal H} = \frac{1}{2}C_{ijkl} E_{ij}E_{kl}$ and curl-free condition
on the electric field coming from the electrostatic limit of the
Faraday law. The latter implies the existence of an electrostatic
potential that plays the role of an effective phonon-like field that,
unlike crystals arises in the absence of spontaneous breaking of
translational symmetry. The rank-4 elastic tensor $C_{ijkl}$ is to be
determined experimentally and is generically heterogeneous and
anisotropic. The formulation then allows an efficient computation of
the stress-stress correlations associated with a distribution,
geometry and topology of the force-chain network via
$\langle E_{ij}E_{kl}\rangle$ correlator. The latter gives 4-fold
pitch-point singularities characteristic of the tensor gauge theory
\cite{PretkoZhaiLRdualityPRB2019, PVCPN_pinch}.

In contrast to a tensor gauge theory of {\em crystalline} solids that,
as we reviewed here can be derived explicitly through duality
\cite{PretkoLRdualityPRL2018}, this gauge theory formulation of
amorphous solids is a conjecture that needs the support of numerics
and experiments.  Indeed the measured averaged stress-stress
correlations are well fit by the electric field correlator of the
vector-charge tensor gauge theory \cite{amorphous_fracton}. 
Fitting this to numerics and experiments \cite{BehringerPRL2001,
  BehringerNature2011} allows one to extract the average
pseudo-dielectric tensor $C_{ijkl}$ that fully characterizes the
emergent static elasticity of the amorphous solid. The resulting
tensor gauge theory can then be used to further explore solid's
phenomenology, such as response to perturbations, melting, and
dynamics.

\subsubsection{Elastic sheets}

Another interesting connection developed in \cite{Moessner}
is the application of fractonic tensor gauge theory to elastic thin
sheets and their associated defects like folds and tears. With this
the authors presented a tensor gauge theory view of the kirigami
mechanics. The fractonic dual theory transparently encodes a number of
known properties of such defects in thin sheets. They showed that the
observation that folding of a sheet of paper can only be done along a
straight line can be interpreted using the language of fractons and
restricted mobility of vector-charge tensor gauge theory.

In more detail, in Ref.\cite{Moessner} the sheet elasticity is
formulated in terms of its out-of-plane flexural field $h(x,y)$,
(questionably) neglecting the in-plane phonon displacements
${\bf u}(x,y)$.  Such model then corresponds to a fluid, rather than
an elastic membrane\cite{JerusalemWS2ndEdLR, LRmembranes}.  The
sheet's local momentum density $\sim \p_0 h$ is identified with the
scalar magnetic flux density $B$, sheet's curvature tensor
$\partial_\alpha\partial_\beta h$ with tensor electric field
$E_{\alpha\beta}$, and flexural modes with a quadratically dispersing
photon.  A ``tear'' defect is characterized by a nonzero closed line
integral of $\partial_\alpha h$ around the end of the defect, thereby
capturing a (non-quantized) discontinuity $\Delta h$ across a
tear. This out-of-plane discontinuity ray has some formal similarity
with the in-plane dislocation defect.  A ``fold'' defect - an
undeformable line along which there is a sheet's tangent vector
discontinuity across the fold, characterized by a closed line integral
of $\partial_\alpha\partial_\beta h$. It maps onto a fractonic vector
charge -- endpoint of the fold -- of the tensor gauge theory.  It is
hoped that such formulation can be useful in exploration of quantum
dynamics and statistical mechanics of kirigami sheets.

\subsubsection{Quasiperiodic systems}

Another interesting example is that of quasicrystals (QC), whose
elasticity-fracton duality was investigated in \cite{Surowka_QC}.  As
developed by its pioneers\cite{KKL_QC, LS_QC, LRT_QC} and discussed in
detail in Ref.\onlinecite{QCreview}, the elasticity of the QCs is
characterized two set of low-energy modes: phonons, described by the
symmetric strain tensor $u_{ij}$, and phasons, described by a general
rank-$2$ tensor $w_{ij}$. Consequently, the equations of motion are
formulated in terms of two stress tensors, $T_{ij}$ and
$H_{ij}$. These can be defined as derivatives of the Lagrangian
density
\be
\la{eq:stressQC} T_{ij} = -\frac{\p \mathcal L}{\p
  u_{ij}}\,, \qquad H_{ij} = -\frac{\p \mathcal L}{\p w_{ij}}\,,
\ee
with, as clear from \eqref{eq:stressQC}, the stress tensor $H_{ij}$,
non-symmetric. Duality transformation follows the steps reviewed in
Sec. \eqref{sec:ElasticDuality}. Under duality each QC stress is
described by a dual tensor gauge field, with a traceless scalar charge
theory $A_{ij}, A_0$ for $T_{ij}$ and the general rank-$2$ tensor
$\mathcal A_{ij}, \mathcal A_0$ characterizing $H_{ij}$.  In the dual
gauge theory these degrees of freedom are coupled and the Lagrangian
takes the Maxwell form, \emph{i.e.}, it is formulated as a quadratic
form in terms of tensor electric and magnetic fields.

Associated with phonons and phasons, QCs exhibit two types of
topological defects: those of $u_{ij}$ (\emph{i.e.}, dislocations and
disclinations) and defects of $w_{ij}$, known as the stacking
faults.\cite{LRT_QC} Disclinations are scalar charges, while
stacking faults are vector charges coupled to $A_{ij}, A_0$ and
$\mathcal A_{ij}, \mathcal A_0$ tensor gauge fields. Mobility of the
dislocations in QCs (and QC SPTs) was carefully studied in \cite{EHPG_QC},
where it was found that the dislocations are lineons with the mobility
direction determined by the Burgers vector \emph{and} additional
topological information.

A particular example of a quasiperiodic system is the Moire
superlattice generated in twisted bilayer graphene
\cite{TBG_QC_frac}. The phasons in Moire systems correspond to
relative displacement of the layers. Singularities of the phason modes
in this context were referred to as discompressions in
\cite{TBG_QC_frac}, which were indeed found to be immobile.

\section{Global symmetries and gauge theories}
\la{GT}

Tensor gauge theories describe fields that naturally mediate
interactions between fractons in a manner similar to electromagnetism.
They were first introduced by Kleinert \cite{Kleinert_dual} in
discussion of dual approach to the melting transition. Later lattice
versions of these gauge theories appeared as low-energy effective
theories of spin liquids \cite{XH_grav}.

We have already encountered such gauge theories in Section
\ref{Defects}, as duals of elastic systems. Here we will present a
symmetry perspective on these theories and explain how they arise from
gauging an abstract symmetry algebra --- \emph{multipole
  algebra}. This approach allows us to generalize tensor gauge
theories to multipole gauge theories that are related to both
anisotropic liquid crystals and fractal surface codes.

\subsection{General symmetric tensor gauge theories}
Tensor gauge theories such as \eqref{eq_gMaxwell} arise from gauging a
global algebra of conserved charge and multipole moments.  Such
algebra has to include spatial symmetries such as translations and
rotations as well because they do not commute with multipole
charges. The simplest case of this algebra includes all multipole
charges up to some rank $r$ and all translations and rotations.

More formally it is described by a set of commutation relations. Let
$T_i$ and $R_{ij}$ be generators of translations and rotations, and
$Q^{(n)}_{i_1i_2\ldots i_n}$ be the charge corresponding to the $n$-th
multipole moment. The multipole algebra $\mathfrak M_{n,k}$ then takes
the form,
\begin{align} 
\la{eq_mult1}
&\left[T_i, T_j \right] = 0\,, \qquad \left[ R_{ij}, T_k \right] = \delta_{k[i} T_{j]}\,,
\\ \la{eq_mult2}
&\left[ R_{ij}, R_{kl} \right] = \delta_{[k[i} R_{j]l]}\,, \quad [Q^{(n)}_{i_1i_2\ldots i_n}, Q^{(m)}_{i_1i_2\ldots i_m}]=0\,,
\\ \la{eq_mult3}
&\left[T_{j}, Q^{(n)}_{i_1i_2\ldots i_n}\right] = \sum_{r=1}^n \delta_{j i_r} Q^{(n-1)}_{i_1i_2\ldots \hat{i}_r\ldots i_{n}}\,\, \quad \forall\; n>k
\\ \la{eq_mult33}
& \left[T_{j}, Q^{(k)}_{i_1i_2\ldots i_k}\right] = 0\\ 
\la{eq_mult4}
&\left[R_{jk}, Q^{(n)}_{i_1i_2\ldots i_n}\right] = \sum_{r=1}^n \delta_{i_r [j} Q^{(n)}_{k] i_1 \ldots \hat{i}_r\ldots i_n}\,,
\end{align}
where $\hat{i}_r$ indicates that the index $i_r$ should be omitted and
$n \geq k$ and the square brackets in Eq.\eqref{eq_mult1} indicate antisymmetrization $\delta_{k[i} T_{j]} =  \delta_{ki} T_{j} -  \delta_{kj} T_{i}$. Eq.\eqref{eq_mult33} indicates that $Q^{(k)}_{i_1i_2\ldots i_k}$ is the fundamental charge that happens to be a rank-$k$ tensor.

Given the conserved charges $Q^{(n)}_{i_1i_2\ldots i_n}$ we postulate
a local conservation law in a form of continuity equation
\be
\la{eq_general_cont_k} \p_0\rho_{i_1 \ldots i_k} + \p_{i_{n-k}}
\ldots \p_{i_{n}} J^{i_1 \ldots i_{n-k} \ldots i_n } = 0\,,
\ee
where $J^{i_1 \ldots i_{n-k} \ldots i_n }$ is a symmetric tensor
current. Eq. \eqref{eq_general_cont_k} implies that a tensor charge
$Q^{(k)}_{i_1\ldots i_k}$ is conserved as well as its first $n-k$
moments. In other words the indivisible unit charge in $\mathfrak M_{n,k}$ is a rank$-$ tensor. 
\begin{align} 
&Q^{(k)}_{i_1 \ldots i_k}  = \int dx \,\,\, \rho_{i_1 \ldots i_k}\,,
\\
&Q^{(m)}_{i_1 \ldots i_m} = \int dx \,\,\, x_{i_{k+1}}\cdot \ldots \cdot x_{i_{m}} \rho_{i_1 \ldots i_k}\,,\quad m\leq n-k\;.
\end{align}

We now introduce a set of gauge fields that are conjugate (sourced by)
to the tensor charge density $\rho_{i_1 \ldots i_k}$ and the tensor
current $J^{i_1 \ldots i_n }$
\be
A_{0,i_1 \ldots  i_{k}}\,, \qquad A_{i_1 \ldots  i_{n}}\,.
\ee
Given these fields we can modify the Lagrangian of a theory that is invariant under the multipole algebra $\mathfrak M_{n,k}$ as follows (this is known as gauging) 
\be
\delta\cL =   A_{0,i_1 \ldots  i_{k}} \rho^{i_1 \ldots  i_{k}}+ A_{i_1 \ldots  i_{n}} J^{i_1 \ldots  i_{n}}\,.
\ee
Then, requiring invariance under  gauge transformations 
\begin{align}
\la{eq_gt1}
&\delta A_{i_1 \ldots  i_n} = \sum_{r=1}^n \p_{i_r} \lambda_{i_1\ldots\hat{i}_r\ldots i_n}\,,
\\ \la{eq_gt2}
&\delta A_{0,i_1 \ldots  i_{k}} = \sum_{r=1}^n \p_0\lambda_{i_1\ldots\hat{i}_r\ldots i_n}\,.  
\end{align}
 enforces the continuity equation \eqref{eq_general_cont_k}.

 Precise structure of \eqref{eq_gt1} - \eqref{eq_gt2} depends on
 $k$. If the lowest conserved moment is the (scalar) charge $Q^{(0)}$,
 \emph{i.e.} $k=0$ the gauge parameter takes general form
\be
\lambda_{i_1\ldots \ldots i_{n-1}} = \p_{i_1} \ldots \p_{i_{n-1}} \lambda\,.
\ee
  If the lowest conserved moment is $Q^{(k)}_{i_1i_2\ldots i_k}$, the algebra still makes sense assuming that the commutator between translations and $Q^{(k)}_{i_1i_2\ldots i_k}$ vanishes $[T_j, Q^{(k)}_{i_1i_2\ldots i_k}]=0$. If that is the case, we say that the theory has a rank-$k$ tensor charge and the higher moments of this charge are conserved. When $k=1$ this is known as vector charge theory.
  
The density and current can then be found by the usual variational prescription
\be\la{eq_rho_J}
\rho_{i_1 \ldots  i_{k}} = \frac{\delta\cL}{\delta A_{0,i_1 \ldots  i_{k}}}\,, \quad J^{i_1 \ldots  i_n } = \frac{\delta\cL}{\delta A_{i_1 \ldots  i_n }}\,.
\ee

The gauge-invariant electric field $E_{i_1 \ldots  i_{n}}$ is easy to construct. It is given by
\be\la{eq:generalE}
E_{i_1 \ldots  i_{n}} = \p_0 A_{i_1 \ldots  i_n} - \sum_{r=1}^n \p_{i_r} A_{0,i_1\ldots\hat{i}_r\ldots i_n}\,.
\ee
The Gauss's law generating \eqref{eq:generalE} takes form
\be
 \p_{i_{k+1}}\ldots \p_{i_{n}} E_{i_1 \ldots  i_{n}}= \rho_{i_1 \ldots  i_{k}}\,.
\ee
Gauge invariant magnetic field can be defined in all of the above
cases, however it explicit form depend on the theory and we do not
provide its general expression.

In \cite{GromovMultipole} the algebra \eqref{eq_mult1}-\eqref{eq_mult4} was referred
to as the maximally symmetric multipole algebra.

\subsection{Examples of symmetric tensor gauge theories}

The \emph{first} example is the scalar charge theory \cite{genemPretko}. It requires the conservation of dipole moment only. The multipole algebra takes form
\begin{align} 
\la{eq_mult1E1}
&\left[T_i, T_j \right] = 0\,, \qquad \left[ R_{ij}, T_k \right] = \delta_{ki} T_{j} - \delta_{kj} T_{i}\,,
\\ \la{eq_mult2E1}
&\left[ R_{ij}, R_{kl} \right] = \delta_{[k[i} R_{j]l]}\,, \quad [Q^{(1)}_{i}, Q^{(1)}_{j}]=0\,,
\\ \la{eq_mult3E1}
&\left[T_{i}, Q^{(1)}_{j}\right] = \delta_{ij}Q\,, \qquad \left[T_{i}, Q\right]  = 0
\\ \la{eq_mult4E1}
&\left[R_{jk}, Q^{(1)}_{i}\right] = \delta_{ki} Q^{(1)}_{j} - \delta_{kj} Q^{(1)}_{i}\,.
\end{align}
A field theory invariant under \eqref{eq_mult1E1}-\eqref{eq_mult4E1} can readily be written. The only degree of freedom is a real scalar $\phi$ with the Lagrangian given by
\be\la{eq_dipole_FT}
\cL = (\p_0\phi)^2 - (\p_i \p_j \phi)^2\,,
\ee
and is the Lifshitz models with ubiquitous
applications \cite{LakeDBHM,SLN_SSB,ShuHengSeiberg,LRLifshitz}
Lagrangian \eqref{eq_dipole_FT} is invariant under a polynomial shift symmetry
\be\la{eq_dipole_cons}
\delta \phi = c_0 + c_i x^i\,,
\ee
which by virtue of Noether's theorem implies the conservation of total charge and total dipole moment. Indeed, there are $d+1$ independent symmetry parameters leading to $d+1$ conserved quantities.

To gauge \eqref{eq_dipole_cons}  we introduce a tensor gauge fields as follows
\be\la{eq_dipole_FTg}
\cL = \frac{1}{2}(\p_0 \phi - A_0)^2 - \frac{1}{2}(\p_i \p_j \phi - A_{ij})^2\,.
\ee
The density and tensor current are then given by \eqref{eq_rho_J} 
\be
\rho = \frac{\delta\cL}{\delta A_0} = \p_0 \phi\,, \qquad J^{ij} = \frac{\delta\cL}{\delta A_{ij}} = \p_i \p_j \phi\,  
\ee
and satisfy the continuity equation
\be
\p_0 \rho + \p_i \p_j J^{ij} = 0\,, 
\ee
which implies the dipole conservation.

Our \emph{second} example is the traceless scalar charge theory. Its symmetry algebra is a sub-algebra of $\mathfrak M_2$ where the conserved quantities are 
\be
Q^{(1)}_i\,,\qquad  \Delta = \tr\left(Q^{(2)}_{ij}\right)\,.
\ee 
The only additional (compared to \eqref{eq_mult1E1}-\eqref{eq_mult4E1}) non-trivial commutation relation is 
\be\la{eq_mult1E2}
[T_i, \Delta] = Q_i\,.
\ee
A Lagrangian invariant under the symmetry algebra \eqref{eq_mult1E1} - \eqref{eq_mult1E2} take form
\be\la{eq_dipole_traceless_FT}
\cL = (\p_0\phi)^2 - \left(\left[\p_i \p_j - \frac{1}{d}\delta_{ij}\p^2\right] \phi\right)^2\,.
\ee
Lagrangian \eqref{eq_dipole_traceless_FT} is invariant under a polynomial shift symmetry
\be\la{eq_dipole_cons2}
\delta \phi = c_0 + c_i x^i + \tilde c x_kx^k\,.
\ee
The last term leads to an extra conservation law for the trace of the quadrupole moment. Gauging leads to the traceless scalar charge theory with a traceless tensor potential, while the tensor current satisfies an extra constraint $\delta_{ij}J^{ij} = 0$.

Finally, we discuss a general scalar charge theory invariant under the multipole algebra $\mathfrak M_{n,0}$. The commutation relations are given by \eqref{eq_mult1}-\eqref{eq_mult4} with $k=0$.
The Lagrangian takes the form
\be
\cL = \frac{1}{2}(\p_0\phi)^2 - \frac{1}{2} (\p_{i_1} \ldots \p_{i_n} \phi)^2\,,
\ee
which is invariant under a general polynomial shift symmetry
\be\la{eq_gen_npole}
\delta \phi = c_0 + \sum_{m=1}^n c_{i_1 \ldots i_m} x^{i_1} \cdot \ldots \cdot x^{i_m}\,.
\ee
The local conservation law that follows from the Noether's theorem takes form \eqref{eq_general_cont_k} with $k=0$. 

To gauge \eqref{eq_gen_npole} we introduce a general rank-$n$ tensor gauge field as follows 
\be
\cL = \frac{1}{2}(\p_0\phi - A_0)^2 - \frac{1}{2} (\p_{i_1} \ldots \p_{i_n} \phi - A_{i_1 \ldots i_n})^2\,.
\ee
The density and tensor current are then given by \eqref{eq_rho_J} with $k=0$.

\subsection{General multipole algebra}
So far we have assumed that the tensor gauge theories discussed above
are invariant under continuous rotations. This does not have to be the
case because we expect (at least) some of those theories to emerge
from UV lattice models and to thereby inherit the lattice
symmetries. We will reduce our discussion to scalar charge theories.

To incorporate the lattice symmetries we observe that every conserved multipole moment is associated to a polynomial $P(x)$. For example, conserved dipole moment is associated to $d$ monomials of degree $1$: $x_1,x_2,\ldots,x_d$. Generally a component of the multipole tensor $Q_\alpha^{(n)}$ is obtained by integrating a polynomial of degree $n$, $P_\alpha^{(n)}(x)$ against the charge density
\be\la{eq_generalmult}
Q_\alpha^{(n)} = \int d x P_\alpha^{(n)}(x) \rho(x)\,.
\ee
 The general index $I$ may transform in an irreducible representation of a point group. We now turn to a more formal description of the general multipole algebra.

To describe the algebra we introduce the general multipole moment as follows. Let $P_\alpha^{(n)}(x)$ be a homogeneous polynomial of degree $n$. Then multipole moment corresponding to $P_\alpha^{(n)}(x)$ is defined be \eqref{eq_generalmult}.

Commutation relations between these multipole moments and spatial symmetries form the multipole algebra
\begin{align}
&\left[T_{\hat{\mathbf{r}}}, Q^{(n)}_\alpha\right] =  f^{(n)}{}_\alpha{}^\beta Q^{(n-1)}_{\beta}\,,
\\ \la{eq_mult_gen}
&\left[R_{\hat{\mathbf{r}}}, Q^{(n)}_\alpha\right] = g^{(n)}{}_\alpha{}^\beta Q^{(n)}_{\beta}\,,
\end{align}
where $T_{\hat{\mathbf{r}}}$ is a translation in direction
$\hat{\mathbf{r}}$, while $R_{\hat{\mathbf{r}}}$ is rotation about
$\hat{\mathbf{r}}$, while $f^{(n)}{}_\alpha{}^\beta$ and
$g^{(n)}{}_\alpha{}^\beta$ are the structure constants. In general,
the rotations can include either a subgroup of $SO(d)$ or be
discrete. For example, in the gauge theory approach to the Haah code
there is an $SO(2)$ rotation symmetry, while
$\hat{\mathbf{r}} \propto (1,1,1)$ \cite{GromovMultipole}.

Given the polynomials we define a set of homogeneous differential operators $D_\alpha$ that annihilate \emph{all} $P^{(n)}_I$ simultaneously
\be\la{eq:derivative}
D_\alpha P^{(n)}_I = 0\quad \forall \,\,\,I,n\,.
\ee
Then the local conservation laws take form
\be\la{eq_consM}
\p_0\rho + \sum_\alpha D^\dag_\alpha J^\alpha = 0\,.
\ee
where $J^\alpha$ are the multipole currents and $D^\dag_\alpha$ is obtained from $D_\alpha$ via integration be parts.

Gauging procedure follows the same logic as in the previous
section. We introduce the gauge fields $A_\alpha$ and $A_0$ conjugate
to the multipole current $J^\alpha$ and density $\rho$. These gauge
fields are labeled by an abstract index (which also can transform in
an irreducible representation of the rotation group), and are neither
$1$-forms no symmetric tensors. A general Lagrangian invariant under
the multipole algebra is supplemented with \be \delta\cL = \rho A_0 +
J^\alpha A_\alpha\,.  \ee The gauge transformation law takes form
\be\la{eq:gaugetr} \delta A_\alpha = D_\alpha \lambda\,, \quad \delta
A_0 = \p_0 \alpha \ee and ensures \eqref{eq_consM}.

Using these $D_\alpha$ we can construct the electric field and the Gauss's law as follows. The electric field is invariant under \eqref{eq:gaugetr} and is given by 
\be
E_\alpha = \p_0 A_\alpha - D_\alpha A_0\,,
\ee
satisfying the Gauss's law with a generalized divergence,
\be
\sum_{\alpha} D^\dag_\alpha E_\alpha = \rho\,.
\ee

\subsection{Relation to the symmetric case}

Multipole algebra includes the symmetric case as a special case. Here we illustrate how it works on the example of traceless scalar charge theory. In $d$ spatial dimensions there are $d+1$ polynomials
\be
P^{(1)}_1(x) = x_1\,,\ldots\,,P^{(1)}_d(x) = x_d\,, \quad P^{(2)}_1(x) = x_i x^i\,.
\ee
The differential operators are of degree $2$. The index $\alpha$ can be represented as a multi-index $\alpha = (i,j)$ with the differential operators given by 
\be
D_{i,j} = \p_i \p_j - \frac{1}{d}\delta_{ij}\p^2\,,
\ee
where it is clear that \eqref{eq:derivative} holds, \emph{i.e.} $D_{i,j}$ annihilates all the polynomials $P^{(1)}_1(x), \ldots, P^{(1)}_d(x),P^{(2)}_1(x)$. This algebra includes all translations and continuous rotations.

\subsection{Gaussian free field with multipole symmetries}

Next we discuss an explicit example of a free field theory that is invariant under a general multipole algebra. Consider a real scalar field $\phi$ and a set of homogeneous polynomials $P^{(n)}_I(x)$. We will construct a Lagrangian invariant under the following transformation
\be\la{eq:Msymmetry}
\phi \longrightarrow \phi + P^{(n)}_I(x)\,.
\ee
To the lowest order in derivatives the most general action takes form
\be\la{eq:multaction}
\cL = \oh (\p_0\phi)^2 + \oh\sum_\alpha (D_\alpha \phi)^2
\ee
by virtue of \eqref{eq:derivative}. The symmetry \eqref{eq:Msymmetry} leads to the conservation of \eqref{eq_generalmult}.

We can also utilize \eqref{eq:multaction} to obtain the multipole gauge theory structure. Indeed, the symmetry \eqref{eq:Msymmetry} can be gauged by replacing derivatives as follows

\be\la{eq:multaction_gauged}
\cL = \oh (\p_0\phi - A_0) - \oh\sum_\alpha (D_\alpha \phi - A_\alpha)^2\,,
\ee
where $\Phi$ is the scalar potential. The action
\eqref{eq:multaction_gauged} is invariant under the gauge
transformation \eqref{eq:gaugetr}. The conserved charges are
explicitly given by
\be
Q^{(n)}_I = \int \p_0\phi P^{(n)}_I\,.
\ee

\subsection{Multipole gauge theory of a smectic}
Next we discuss a simple example of multipole gauge theory that
arrises as a dual theory to elasticity of a quantum smectic
phase \cite{LRsmecticPRL2020, Z3LRsineGordonPRB}. We will show that a 2+1d
quantum smectic is dual to a multipole gauge theory. We start with the
following Lagrangian density \cite{deGennesProst} that describes a smectic phase
at long distances
\be
\cL = \oh {\p_0 u}^2 -
\oh\kappa(\pt_y u)^2 - \oh\kappa(\lambda\pt_x^2 u)^2\,,
\ee
where the layers are perpendicular to the $y$ axis and extend
along the $x$ axis. We remind the reader that a smectic is a liquid
crystal phase that 
spontaneously breaks rotational symmetry (by a choice of layers'
orientation) and one out of the two translation symmetries. It can be
viewed as a periodic array of 1d liquids with a period of order $\lambda$ along
$y$ axis. We denote
\be D_1^\dag = \p_w\,,
\qquad D_2^\dag = \lambda\p_u^2\,.
\ee
In terms of these derivatives
the action is
\be
\cL = \frac{1}{2} {\p_0 u}^2 -
\oh\kappa (D_1^\dag u)^2 - \oh\kappa(D_2^\dag  u)^2\,.
\ee
We introduce auxiliary variables using Hubbard-Stratonovich trick
\be
\cL =  P \dot u -
\frac{P^2}{2} - (D_1^\dag u) T_1 - (D_2^\dag u) T_2
+\epsilon\frac{T_1^2}{2}+ \epsilon\frac{T_2^2}{ 2}
\ee
Integrating out
the phonon $ u$ we find a constraint
\be \p_0 P - D_1 T_1 - D_2
T_2 = 0\,.  \ee This equation is solved by \bea &&T_I =
\epsilon_{IJ}(\p_0A_J - D_J A_0) = \epsilon_{IJ}E_J\,,
\\
&&P = B = \epsilon_{IJ}D_IA_J\,,
\eea
where $\epsilon_{IJ}$ is the
Levi-Civita symbol and $I,J = 1,2$. The gauge redundancy of the solution is

\be\la{eq:2DMGT} 
\delta A_I = D_I\alpha\,, \quad \delta A_0 =
\dot\alpha \,,
\ee 
which is exactly a multipole gauge theory
structure. The Gauss's law that generates \eqref{eq:2DMGT} is given by
\be\la{eq:2DGauss} D_I^\dag E_I =\rho\,.  \ee The defect density
$\rho$ is the density of smectic disclinations. The defect matter
conserves the dipole moment in $u$ direction, which can be seen
directly from \eqref{eq:2DGauss}. Disclination dipole extended in the
$u$ direction is a dislocation with the Burgers vector in $w$
direction. The dislocations are completely mobile
\cite{kleman2007soft}, whereas the disclinations are $1$D particles
(also known as lineons) that can only move in the $w$ direction. The
low energy phonon is described by the multipole gauge theory with the
generalized Maxwell action
\be\la{eq:2DMGT2}
\cL = \frac{1}{2}\epsilon(E_1^2 + E_2^2) - \oh B^2\,.
\ee
As required $\cL$ admits is a single smectic mode with a linear dispersion along
the $y$ axis and quadratic dispersion along the $x$ axis.

\subsection{$U(1)$ Haah code in three dimensions} 

Next we turn to a discussion of the ``$U(1)$ Haah code'' studied in
\cite{BB_fractal,GromovMultipole}.
We begin by postulating the symmetries
\be\la{eq:Haah}
\delta \phi = c_0 + c_{1} P_1^{(1)} +  c_{2} P_2^{(1)} + c_{3} P_1^{(2)} +  + c_{4} P^{(2)}_2\,, 
\ee
where
\bea
&&P^{(1)}_{1} = x_1-x_2\,,\quad P^{(1)}_{2}= x_1+x_2-2x_3\,,
\\
&&P^{(2)}_{1} = (x_1-x_2)(x_1+x_2-2x_3)\,,
\\
&& P^{(2)}_{2} =(2x_1-x_2-x_3)(x_2-x_3)\,.
\eea

\begin{figure*}
\includegraphics[width=\textwidth]{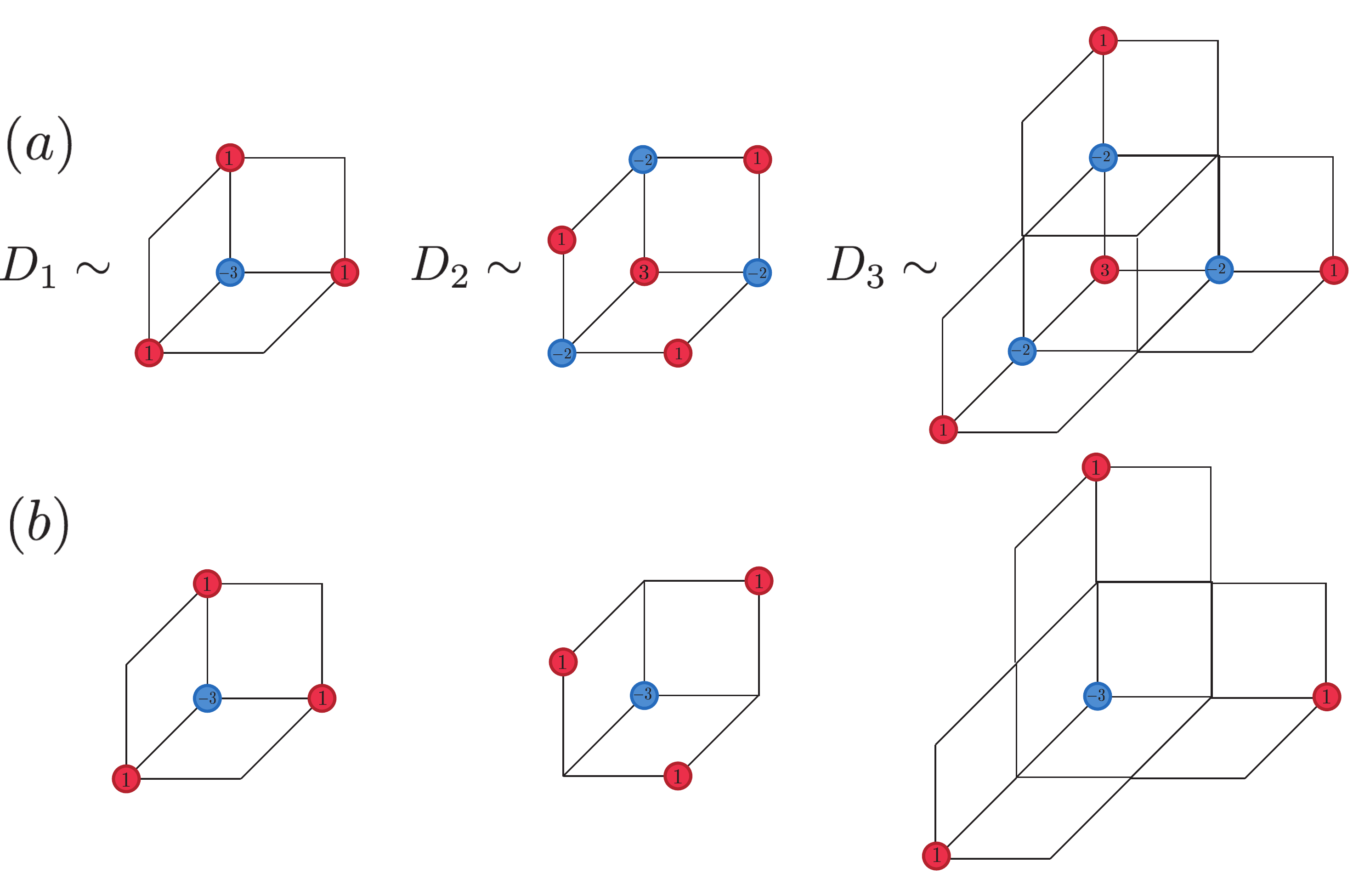}
\caption{(a) The elementary charge configurations, corresponding to the invariant derivatives $D_I$, for the effective theory for the $U(1)$ Haah code \eqref{eq:HaahDerivatives} charge configurations. These charge configurations violate conservation of the dipole moment in $(1,1,1)$ direction. (b) A different basis of elementary charge configurations. The first two configurations are precisely the ones studied in [\onlinecite{BB_fractal}], while the last charge configuration is allowed by symmetries and is linearly independent from others.}
\label{Haah3D}
\end{figure*}

These polynomials have been chosen after examining the elementary fracton configurations in the Haah code \cite{GromovMultipole} and generalizing them to the $U(1)$ conserved charge. These configurations carry dipole moment in the $(1,1,1)$ direction, which leads us to enforce conservation of the dipole moment in the $(111)$ plane. Polynomials of the degree $2$ were chosen in a similar manner. We will also enforce the $SO(2)$ symmetry  in the $(111)$ plane. This leaves us with \emph{three} invariant derivatives that take form
\be\la{eq:HaahDerivatives}
D_1 = q^i \p_i\,, \quad D_2 =  q^{ij}_1 \p_i\p_j\,, \quad  D_3 =  q^{ij}_2 \p_i\p_j\,,
\ee
where
\bea\nonumber
q^{i} =\left(
\begin{array}{ccc}
 1  \\
 1  \\
 1 \\
\end{array}
\right),
\,\,\,
q^{ij}_1 =\left(
\begin{array}{ccc}
 1 & 0 & 0 \\
 0 & 1 & 0 \\
 0 & 0 & 1 \\
\end{array}
\right),
\,\,\,
q^{ij}_2 =\left(
\begin{array}{ccc}
 0 & \frac{1}{2} & \frac{1}{2} \\
 \frac{1}{2} & 0 & \frac{1}{2} \\
 \frac{1}{2} & \frac{1}{2} & 0 \\
\end{array}
\right)\,.
\eea

We then gauge these symmetries, as explained in the previous sections. The Lagrangian describing dynamics of the gauge field is given by
\be \la{eq:HaahLag}
\cL =  \sum_I E_I^2 - B_1^2 - B_2^2\,,
\ee 
Where the magnetic fields are  given by 
\be
B_{I} = \epsilon_{IJK} D_J A_K\,,
\ee
and the Gauss's law takes form
\be\la{eq:GaussHaah}
\sum_\beta D^\dag_\beta E_\beta = \rho\,.
\ee

The $U(1)$ Haah model has a hidden infinite symmetry. We need to introduce a bit of notation. First we define a basis $\mu^1_i, \mu^2_i$ in the plane where the dipole moment is conserved. One choice is ${\bf \mu}^1 = (1,-1,0)$ and ${\bf \mu}^2 = (1,1,-2)$. With this basis at hand we introduce new variables $\mathrm{x}=\mu^{1}_i x^i/|\mu^1|$ and $\mathrm{y}=\mu^{2}_i x^i/|\mu^2|$. Then all invariant derivatives $D_\alpha$ (and, consequently the Lagrangian) are also invariant under an infinite symmetry 
\be\la{eq:conformal}
\delta \phi(\mathrm{z},\bar{\mathrm{z}}, x_3) = f(\mathrm{z}) + g(\bar{\mathrm{z}})\,, \quad \mathrm{z}=\mathrm{x}+i\mathrm{y}\,,
\ee
where $f(\mathrm{z})$ is holomorphic and $g(\bar{\mathrm{z}})$ is anti-holomorphic. This is an example of a well-known ``sliding'' symmetry \cite{barci2002theory}, that appears in physics of smectics \cite{o1998nonlinear} and it can be understood as a continuous version of sub-systems symmetries. 
Finally, the $U(1)$ Haah model exhibits an anisotropic scaling symmetry, which takes form
\be
t\rightarrow \lambda t\,, \,\,\, \mathrm{x}\rightarrow \lambda^{\frac{1}{2}} \mathrm{x}\,,\,\,\, \mathrm{y}\rightarrow\lambda^{\frac{1}{2}}\mathrm{y}\,,\,\,\, x_3 \rightarrow \lambda x_3\,,\,\,\,\phi\rightarrow \lambda^{-\frac{1}{2}}\phi\,. 
\ee
It was shown in \cite{Gromov_smectic} that gauge theory for $U(1)$ Haah code
\eqref{eq:HaahLag} is dual to the smectic-A phase in $3$D.

\subsection{Subsystem symmetry}
We now turn to an even more exotic class of much larger, extensive
symmetries -- the so-called \emph{subsystem} symmetries -- that lead to
restricted mobility and multipole gauge theories upon
gauging \cite{fracton1, fracton2}. These symmetries were initially
defined on a lattice for various spin models, but, as we illustrate
can be extended to the continuum.

While a covariant theory of subsystem symmetries has not yet been
developed, we can understand a class of these symmetries as an
infinite-dimensional generalization of the multipole algebra discussed
in the previous subsection. In our development we assume that the
physical model is defined on a flat space and that we are given a set
of lines, planes, or hyperplanes that foliate the space.

As a  pedagogical example we consider a model of a real
scalar field in 2d, $\phi(x_1,x_2)$, with the following
transformation as an example of subsystem symmetry,
\be\la{eq:sss}
\delta \phi = f_1(x_1) + f_2(x_2)\,,
\ee
where $f_1(x_1)$ and $f_2(x_2)$ are \emph{arbitrary} functions of
$x_1$ and $x_2$. The set of lines consists of two families: (i)
lines parallel to $x_2$ and (ii) lines parallel to $x_1$. Any 2d
lattice provides enough structure to develop a set of subsystem
symmetries.

The algebra of subsystem symmetries is infinite-dimensional.  Its
action is intermediate between a global symmetry, that acts on full
d-dimensional space and gauge redundancy that acts on individual
sites. We can also interpret it as an infinite-dimensional
generalization of the multipole algebra by representing the functions
$f_1, f_2$ as Taylor series in $x_1,x_2$, correspondingly \be \delta
\phi = \sum_{n\geq 0} c_n x_1^n + \sum_{m\geq 0} b_m x_2^m\,, \ee
where $c_n$ and $b_m$ are arbitrary coefficients. Viewed this way the
symmetry implies conservation of arbitrary high multipole moments in
one of the axes \be Q^{(n)}_{11\ldots 1}\,, \qquad Q^{(n)}_{22\ldots
  2}\,, \quad n\geq0 \ee The Lagrangian invariant under \eqref{eq:sss}
break rotational symmetry down to a discrete subgroup, $C_4$ \be \cL =
\oh (\p_0\phi)2 -\oh (\p_1 \p_2 \phi)^2 + \ldots, \ee and was
analyzed in great detail in Ref. \onlinecite{ShuHengSeiberg}.

Subsystem symmetries can also be gauged. Indeed gauging \eqref{eq:sss}
requires a single ``vector'' potential $A_{12}$ that transforms as
$\delta A_{12} = \p_1 \p_2 \lambda$. There is a single electric field
given by $E_{12} = \p_0A_{12} - \p_1\p_2 A_0$ and the Gauss's law
takes the form,
\be
\p_1 \p_2 E_{12} = \rho\,.
\ee
Other examples of subsystem symmetries are discussed in \cite{GLN_hydro}.

The conservation law following from \eqref{eq:sss}  is encoded in a continuity equation
\be\la{eq:ssc}
\p_0\rho + \p_1 \p_2 J = 0\,,
\ee
where $J$ is the current. It exhibits an infinite number of conserved charges
\be
Q_{1x} = \int_{x_1=x} dx_2 \rho(x_1,x_2)\,,\qquad Q_{2y} = \int_{x_2=y} dx_1 \rho(x_1,x_2)\,,
\ee
which are conserved independently on any line $x$ and $y$,
respectively.  It is thus clear that this conservation law makes
particles that are charged under 
 both $Q_{1x}$ and $Q_{2y}$ completely immobile, while the
dipoles can move perpendicular to their dipole moments.

\subsection{Fracton hydrodynamics}
Conservation laws \eqref{eq_elast} and \eqref{eq:ssc} can, in
principle, be either microscopic or emergent. In either case, the
long-wave, long-time phenomenology is affected if these conservation
laws are present. These effects manifest themselves in the transport
of charge and momentum, with the simplest manifestation of
sub-diffusion.

We illustrate the emergence of sub-diffusion in the simplest case
of a conserved dipole moment. As discussed in earlier sections,
the corresponding charge continuity equation takes the form \eqref{eq_elast},
\be
\p_0\rho + \p_i \p_j J^{ij} = 0\,.
\ee
To describe diffusion of charge we relate the dipole current to the charge density. In equilibrium the dipole current must vanish. Consequently, the constitutive relation between $\rho$ and $J^{ij}$ takes form
\be
J^{ij} = \chi^{-1}\p^i \p^j \rho + \ldots\,,
\ee
where $\chi$ is the susceptibility. The (sub-)diffusion equation then takes form
\be\la{eq:sub-dif}
\p_0\rho + \chi \p^4 \rho = 0\,.
\ee
Eq.\eqref{eq:sub-dif} implies that density perturbation at wavelength
$\lambda$ decays at a characteristic time $\tau \sim
\chi\lambda^4$\cite{GLN_hydro,PretkoLRdualityPRL2018,RHvectorPRL2020,LRsmecticPRL2020} for long wavelengths parametrically far slower than the conventional
diffusive time $\sim\lambda^2$. Such slow relaxation time enhancement has been
observed in cold atomic gasses in tilted optical lattices \cite{subdiffusionPrinceton}. Subdiffusion also emerges in random unitary circuits.\cite{Iaconis_anomalous, Feldmeier_anomalous, Moudgalya_spectral}

Such sub-diffusion straightforwardly generalizes to the case of
conservation of the $n$-th multipole moment, where it gives,
\be
\p_0 \rho + \chi \p^{2+2n} \rho = 0\,,
\ee
leading to even slower characteristic time $\tau \sim \chi\lambda^{2+2n}$.

Another interesting effect appears when subsystem symmetry constrains
the diffusion equation. Consider the charge conservation equation
\eqref{eq:ssc} with dipole symmetry. Relating the generalized current
to the density with the subsystem constraint, we obtain the (sub-)diffusion equation
\be\la{eq:diffSS}
\p_0 \rho + \chi^{-1} \p_x^2 \p_y^2 \rho = 0\,.
\ee
Thus, the subsystem symmetry breaks the rotational symmetry down to a
discrete subgroup of the lattice, with these effects
persisting at longest scales. This is in stark contrast with the
classic diffusion, that has an emergent rotational symmetry to the
lowest order in derivatives. 

%
%
%
%
%
%
%

\section{Conclusions}

\subsection{Summary}

In this Colloquium we have reviewed theoretically inspired, burgeoning
subject of fractonic matter. We began with a model-independent,
symmetry- and conservation-based formulation of fractons -- excitations
with restricted mobility arising in a broad class of exotic models.

The central focus of this review is on the emergence of fractonic
order from elasticity-gauge duality of a broad class of quantum
elasticity models, that include quantum commensurate and
incommensurate supersolid crystals, smectic liquid crystals, hexatic fluids,
amorphous solids, quasi-crystals and elastic membranes, all encoding
some form of multipolar global symmetries. We also discussed a vortex
crystal and a vortex liquid that dualize to a parity-breaking gauge-dual
variants. Building on the familiar boson-vortex duality, we explicitly
reviewed how such dualities lead to interesting tensor and
coupled-vector gauge theories that exhibit fractonic charges, dipoles
and higher multipoles as duals of elastic topological defects, with
gauge fields encoding gapless phonons.

As we discussed, such elasticity-gauge duality is a powerful tool for
discovery of new class of fractonic models. The resulting models can
then also be generalized into a broader class, beyond any elastic dual
connection.  A complementary motivation for the duality studies is
that they provide efficient formulation of the quantum elasticity and
the topological defects dynamics. One striking example is the
prediction of the zero-temperature immobility of disclinations in a 2d
crystal (previously unknown despite decades of its studies in the
elasticity context), that arose purely through this fractonic gauge
theory connection. In the review, we furthermore demonstrated that
gauge duals provide the first field-theoretic formulation of quantum
melting of a crystal and a smectic through a generalized Higgs
mechanism associated with a condensation of dislocations and
disclinations. 

As we discussed, tensor gauge theories can also be studied in
arbitrary dimension and without any relation to elasticity.  We have
presented a general construction of a large class of such gauge
theories based on gauging the \emph{multipole algebra} --- an algebra
of spatial symmetries that includes dipole and higher multipole
symmetries. The resulting multipole gauge theories include the models
obtained from dualities as special cases. We expect that these
theories can serve as templates for identifying exotic gapless
excitations in spin liquids.

We have also discussed an even more exotic subsystem symmetries and
illustrated a formal relation between these symmetries and
infinite-dimensional multipole algebras. Finally, we concluded the
review with a description of the long-time subdiffusive hydrodynamic
of fractonic matter that emerges as a result of multipole conservation
laws.

\subsection{Open problems}

While the field of fractons and associated tensor/multipole gauge
theories has seen a rapid growth in the last ten years, it remains in
an early stage of development, with many open theoretical and
experimental questions.

\subsubsection{Mathematical structure}
Most fundamentally, the mathematical structure of tensor gauge
theories is still not well formulated, currently unclear what replaces
the $G$-bundle of the traditional gauge theories. Consequently, the
topology of the space of tensor gauge fields is poorly
understood. Namely, since the geometric interpretation of tensor gauge
fields is unclear, it is not known how to construct topological
invariants that generalize the Chern numbers. In fact it is certain
that such invariants cannot be purely topological, because any naive
generalization of the Chern number will depend on the spatial metric,
leading to the metric dependence of the ``topological invariant''.
One can also generalize the Chern-Simons theory to the higher rank
case, but the dependence on the metric appears to be unavoidable.
Furthermore, in all cases the inclusion of the metric is in tension
with gauge invariance, finding that the magnetic field is only gauge
invariant in flat space (or on Einstein manifold in certain cases)
\cite{GromovDualityPRL2019, slagle_curved, jensen_curved, hartong_curved}.  It is hoped, however,
that coupled vector gauge theory formulation \cite{RHvectorPRL2020,
  LRsmecticPRL2020} may be more suitable for addressing these
questions.


As we have discussed in the Introduction, fractonic gauge theories
concisely encode quasi-particle restricted mobility and Gaussian
fluctuations. However, these continuum field theories preclude an
encoding of their expected exponential in-system-size ground-state
degeneracy\footnote{Unless lattice regularization is introduced \cite{field, SeibergShao1, SeibergShao2}}, and nontrivial quasi-particle quantum
statistics\cite{QRH_fracton} -- this contrasts strongly with discrete
qubit models (e.g., X-cube), where these properties are well defined
and have been calculated \cite{MaHermele, BB_Higgs}.

As was recently demonstrated \cite{DG_vortices, DMNS_APD}, a gauge
structure and phenomenology similar to that of fractonic tensor gauge
theories also arises in superfluids and the fractional quantum Hall
effect. In fact, a similar form and UV/IR mixing characteristic of
fractonic gauge theories,\cite{ShuHengSeiberg} arise in the
non-commutative gauge theories as well as non-commutative matter
theories coupled to a gauge field. A development and illumination of
these relations remains an open problem.



\subsubsection{Quantum melting and insights on elasticity}
Focussing on specific models and their phenomenology discussed in this
review, we have seen that these dual gauge theories are also of
interest because they provide a formulation of crystal-to-hexatic,
crystal-to-smectic, smectic-to-nematic quantum melting transitions as
generalized Higgs transition associated with a condensation of the
topological defects.  However, these have so far only been analyzed at
the mean-field level thus leaving their true criticality to future
studies.

\subsubsection{Generalizations beyond bosonic elasticity}
We also note that in all the elastic models reviewed here and studied
in the literature, the focus has been on the simplest bosonic
realizations. These lead to bosonic statistics of dislocation and
disclination defects (i.e., bosonic fracton matter) and superfluidity
when crystalline order is lost.  It is natural to study extensions of
the bosonic duality to that of fermions and anyons and to explore the
possibility of nontrivial statistcs of topological defects.

\subsubsection{Experimental void}
Of course the field's most vexing challenge is that of experimental
realizations, that so far have been sorely absent.  In part this is
due to the fact that discrete qubit (e.g., the simplest X-cube)
models, are typically formulated in terms of many-spin commuting
projectors, that are therefore extremely difficult to implement.  A
promising alternative direction is that of the $U(1)$ tensor and
coupled vector gauge theories that have been the focus of this
review. These are related to discrete models through condensation of
higher charge matter \cite{BB_Higgs,MaHermele}. In fact, the 2+1d
crystal-gauge duality strikingly demonstrates a concrete physical
realization of fractonic order in a familiar quantum crystal and other
elastic media.

In tensor-gauge theories, realization-independent quadrupolar
pinch-point singularities have been predicted\cite{PVCPN_pinch,
  rahul2,rahul3, amorphous_fracton} (extending neutron scattering
predictions for conventional spin-liquids in frustrated
magnets\cite{SavaryBalentsQLreview}) and in fact observed in granular
solids.\cite{jam_review} However, so far no smoking gun restricted
mobility experiment has been conceived even in these physical
instantiations of an in-principle fractonic elastic systems. The
immobility of disclinations is not a particularly impressive
observation, especially deep inside a crystal phase, where even
vacancies and interstitials are energetically immobile and behave
classically. Perhaps a study of a He$_4$ crystalline film or 2d
lattices of ultra cold bosons close to a low-temperature quantum
melting transition are good experimental platforms to explore.

Another obstacle in this direction is that in 2d, a disclination
energy is extensively large and thus is not an excitation that can be
easily explored. In contrast, the topology of a spherical crystals,
such as, for example, the Buckminsterfullerene C$_{60}$ molecule and
many other closed structures, as large as C$_{960}$, by Gauss-Bonnet
Theorem automatically ensures $12$ disclinations in the ground state,
whose lattice hopping dynamics can perhaps be studied experimentally.
Extension of dual gauge theories to a spherical geometry with a
detailed analysis remains an problems, and corresponding experiments
is an interesting direction to pursue.

Some further proposals for experimental realization of fractons or tensor gauge fields have appeared in spin liquids \cite{yan_rank2} and Rydberg atoms \cite{xu_fracton, ashvin_fracton}.

This plethora of exciting open problems bodes well for a bright future
of the exciting field of fractons, the current glimpse of which we
presented in this Colloquium.

\section{Acknowledgments}
LR acknowledges support by the Simons Investigator Award from the
Simons Foundation and thanks Michael Pretko, Mike Hermele, Marvin Qi
and Zhengzheng Zhai for collaborations. AG is supported by the NSF
CAREER grant DMR-2045181, the Sloan Foundation, and by the Laboratory
for Physical Sciences through the Condensed Matter Theory Center.


\begin{thebibliography}{99}
\bibitem{mLifshitz} A generalized $m$-Lifshitz model, has elasticity
 with $m$ ``soft'' (Laplacian) axes and $d-m$ complementary ``hard''
 (gradient) axes. In this nomenclature, the conventional classical 3d
 smectic is the $m=2$-Lifshitz model, and 2+1d quantum smectic and 3d
 classical columnar liquid crystal are described by the
 $m=1$-Lifshitz model.  Other generalizations include a nonscalar
  Goldstone mode field as in e.g., tethered
  membranes\cite{LRmembranes, RTtubulePRL, RTtubulePRE}
  and nematic elastomers \cite{XRelastomers}.
\end{thebibliography}

\end{document}